\documentstyle[aps,multicol,epsf]{revtex}
\newcommand{\beq}{\begin{equation}} \newcommand{\eeq}{\end{equation}}
 \newcommand{\al}{\alpha}
\newcommand{\ds}{\displaystyle}

\begin{document}

\title{Mechanisms of Granular Spontaneous  Stratification and Segregation in 
Two-Dimensional Silos}

\author{Pierre Cizeau, Hern\'an A. Makse,\footnote{Present address:
Schlumberger-Doll Research, Old Quarry Road, Ridgefield, CT 06877}
and H. Eugene Stanley}

\address{Center for Polymer Studies and Department of Physics\\ 
Boston University, Boston, MA 02215 USA}

\date{submitted to Phys. Rev. E}

\maketitle

\begin{abstract} 
Spontaneous stratification of granular mixtures has been reported by
Makse {\it et al. } [Nature {\bf 386}, 379 (1997)] when a mixture of
grains differing in size and shape is poured in a quasi-two-dimensional
heap. We study this phenomenon using two different approaches. First, we
introduce a cellular automaton model that illustrates clearly the
physical mechanism; the model displays stratification whenever the large
grains are rougher than the small grains, in agreement with the
experiments. Moreover, the dynamics are close to those of the
experiments, where the layers are built through a ``kink'' at which the
rolling grains are stopped. Second, we develop a continuum approach,
based on a recently introduced set of coupled equations for surface
flows of granular mixtures that allows us to make quantitative
predictions for relevant quantities. This approach includes {\it
amplification\/} (i.e., static grains entering the flow of rolling
grains), a phenomenon neglected in the cellular automaton model.  We
study the continuum model in two limit regimes: the large flux or thick
flow regime, where the percolation effect (i.e., segregation of the
rolling grains in the flow) is important, and the small flux or thin
flow regime, where all the rolling grains are in contact with the
surface of the sandpile. (1) In the thick flow regime, where most
experiments are carried out, the flowing grains are segregated in the
rolling phase; as they are flowing down, the large rolling grains are
convected to the top of the layer, and only the small rolling grains
interact with the sandpile. We include this effect in the continuum
model and find results very close to the experiments. (2) In the thin
flow regime, we find interesting results that are close to the thick
flow limit. However, due to the presence of cross-amplification
processes, we find a small regime that shows stratification when the
small grains are slightly rougher than the large grains, but
stratification is much more pronounced if the large grains are rougher.
We study in detail the dynamical process for stratification, where the
layers are built through a ``kink'' mechanism, and find the dependence
of the size of the layers on the parameters of the system. We find that
the wavelength of the layers behaves linearly with the flux of grains.
We also find a crossover behavior of the wavelength at the transition
from the thin flow to the thick flow regime. We obtain analytical 
predictions for the shape of the kink giving rise to stratification 
as well as the profile of the rolling and static species when segregation
of the species is observed.
Finally, 
we compare our approach to
previous ones.
\end{abstract}



\begin{multicols}{2}
\narrowtext

\section{Introduction}

Segregation is a prominent example of the peculiar properties of
granular matter
\cite{bagnold1,nagel,bideau,mehta,wolf,review5,review6,varenna,hans0}.
For example, shaking a container filled with two types of grains of
different sizes leads not to mixing---as in liquids---but to
segregation, with the large grains on the top of the container and the
small grains on the bottom. This striking behavior and the importance of
mixing problems from a technological point of view have led to a broad
interest in granular materials in the physics community.

Several types of segregation have been investigated, namely,
segregation by vibrating a container (the ``Brazil nut effect'')
\cite{williams,rosato,herrmann,knight,warr}, and segregation in
rotating cylinders, where the segregation occurs through surface flow
\cite{omaka,zik,kaka1,duran,thin2}.

Segregation can also been obtained in the absence of any periodic
oscillation by simply pouring a mixture of grains of different sizes
onto a pile. One experimental set-up, which has attracted much recent
attention, consists of a quasi-two dimensional cell or vertical
Hele-Shaw cell where a mixture of grains is constantly poured next to
one end of the cell [Fig.~\ref{cell}(a)]. When a mixture of small and large
grains is poured into the cell, a pile builds up and the small grains
are observed to segregate near the top of the pile and the large ones
near the bottom of the pile
\cite{brown,bagnold0,williams63,drahun,fayed,savage,meakin}. 
This segregation is due to the different grain sizes, because large
grains roll down more easily on top of small grains than small grains on
top of large grains.

It has been recently observed \cite{makse,makse2,yan,kaka2} that if the
shapes of the species are sufficiently different, a novel type of
segregation can take place: when such a mixture of small rounded grains
and large rough grains is poured between two vertical slabs, a
spontaneous stratification of the mixture in alternating layers of small
and large grains parallel to the pile surface is observed. Additionally,
there is an overall tendency for the large and small grains to segregate
into different regions of the cell. Spontaneous stratification is also
found for more than two species \cite{makse}. The layers are then
ordered forming a sequence parallel to the surface pile
 (from bottom to top, small-medium-large,
small-medium-large for 3 species).

This phenomenon could possibly be relevant in different fields, such
as geomorphology \cite{israel,fineberg}. For example, stones coming
from sand dune solidification (sandstone) present successive
alternation of layers of different types of grains. This regularity
cannot be explained by periodic sedimentation. But one could imagine
that the sand dune was built with sand brought by the wind and flowing
down the slip-face of the dune \cite{bagnold1}, where stratification
could appear as in the experiments of \cite{makse}. This phenomenon
may also have important consequences in industry where processing and
transport of initially mixed grains could lead to stratification.

Spontaneous stratification has been observed in \cite{makse} after a
transient regime and only when the large grains are {\it rougher\/} than
the small grains. 
When the large grains are the smoothest
stratification does not occur but only the segregation 
of the mixture in different regions of the cell occurs, with the
large and rounded being at the bottom of the cell, and the small and rough
at the top.
These phenomena can be qualitatively understood as the
growth of an instability due to the competition of two opposite effects.
(i) On the one hand, if the grains have the same shape (spherical for
example) or, more precisely, the same repose angle, then the large
grains will roll more easily on the top of the small ones than the
reverse. This will lead to segregation, with the small grains at the top
of the pile and the large at the bottom. (ii) On the other hand, if the
grains have the same size but different shape, the rougher grains will
have a larger friction coefficient than the other grains, and we expect
segregation with the rougher grains at the top of the sandpile. In the
case of large grains rougher than small grains, those two effects will
compete, giving rise to an instability, and then to spontaneous
stratification. However, in the case where the smaller grains are
rougher than the large grains, the two effects will contribute in the
same way, and the segregation pattern will be stable, without the
occurrence of stratification \cite{makse3}.

In this paper, we study granular flow in the spirit of
\cite{bouchaud,degennes-french,degennes,mcs}, where one assumes a
sharp distinction between the rolling or fluid phase composed of the
layer of grains flowing down and the static phase or bulk composed of
grains forming the sandpile. The essential feature to understand
sandpile formation is then to describe how the rolling grains are
flowing down and how they interact with the static grains. In this
context, it is important to distinguish two different regimes, the thin
flow regime and the thick flow regime, where the interaction between
rolling grains and static grains are very different. Moreover,
experiments have been done in both regimes, so that both regimes are
important to study.

Here we propose a mechanism to understand spontaneous stratification and
segregation in two-dimensional silos. We adopt two different approaches.
First we introduce a cellular automaton model to give insight into the
proposed mechanism. This model displays stratification as soon as the
large grains are rougher than the small ones, in agreement with
experiments \cite{makse}. The dynamics are also very close to the
experiments \cite{makse}, where the layers are built through a ``kink''
going uphill where the rolling grains are stopped.

Second, we study a continuum model which is built on a phenomenological
formalism describing granular flow on sandpiles. This formalism was
introduced by Bouchaud {\it et al.} \cite{bouchaud} for the case of a
single species, and recently generalized by Boutreux and de Gennes (BdG)
\cite{degennes} to the case of a mixture of two species. The essential
feature to understand sandpile formation is to describe the interaction
between the rolling or fluid phase and the bulk, as well as any
segregation which may occur in the rolling phase itself. In contrast to
the cellular automaton model, this formalism includes {\it
amplification\/} processes, i.e., the conversion of static grains at the
surface of the sandpile into rolling grains by the flow (this feature is
essential to understand avalanching when one tilts a sandpile above the
maximum angle of repose, or to understand avalanche dynamics in rotating
drums).

In the large flux regime, the rolling grains form a thick phase and the
grains are kinematically segregated in the rolling phase, an effect
called kinematic sieving, free surface segregation, or percolation 
[Fig. \ref{cell}(b)]
\cite{drahun,savage,makse2}. Due to this phenomenon the large grains
in the rolling phase are found to rise to the top of the rolling phase
while the small grains sink downward through the gaps left by the motion
of larger grains in the rolling phase. Thus, small rolling grains form
a sub-layer underneath the sub-layer of large rolling grains. Then only
the lower grains of this layer interact with the sandpile. Thus, a
segregation effect that we refer to as percolation takes place inside
the rolling phase where the large grains are convected along the flow to
the top of the layer. Thus, if small rolling grains are present, they
will preferentially be on the bottom of the layer, preventing the large
grains from interacting with the sandpile.

We will develop a continuum formalism that includes the percolation effect,
and show that stratification arises in a similar way as in the
experiments. When the large grains are rougher, stratification is made
of layers spreading all along the sandpile; the layers of the two types
of grains being of the same size for an equal volume mixture of grains.
When the small grains are rougher than the large grains, we observe the
complete segregation of the mixture but not stratification, with the
small and roughest grains being found at the top of the pile in
agreement with the experiments. A thick rolling phase is a condition
for the percolation effect to take place. However a large difference in
size is also necessary to observe the percolation effect: we expect
$d_2/d_1 \stackrel{>}{\sim} 1.5$,
where $d_1$ and $d_2$ are the typical
size of the small and large grains, respectively.

In the small flow regime
each rolling grain is always in contact with the
sandpile and so interacts with it. In this case a thin rolling phase is
expected--- of the order $\approx d_2$--- 
and percolation effects are expected to be irrelevant.  However
strong segregation effects are expected, as in the case of percolation,
since we will consider the case of large size difference between the
grains. The large grains are not being captured on top of static small
grains (although they interact with them) because of the large
difference in size, while the small grains are easily captured on top of
the large static grains. Using the theoretical formalism adapted to
this case, we will show that stratification is observed when the large
grains are rougher than the small grains, in agreement with the
experiments.

Cross-amplification processes---whose influence was greatly reduced in
the case of thick flows due to percolation effects---now plays an
important role since all grains are interacting with the pile and
cross-interactions of the type small/large grains are expected to occur
at the fluid-bulk interface. Due to cross-amplification, the transition
from stratification to segregation does not occur sharply when the small
grains become the roughest. We find a regime where the small grains are
slightly rougher than the large grains, which also gives rise to
stratification. When the small grains are more rounded than the large
grains, then we find the segregation of the grains in agreement with
experiments.

In general, we find that the thick flow regime and the thin flow regime
show similar results regarding the stratification and segregation of the
species. This is valid provided that the grains differ appreciably in
size (in term of the size ratio we expect $d_2/d_1 \stackrel{>}{\sim}
1.5$ \cite{makse2,yan}). Then we study analytically a mechanism for
stratification and segregation valid for both regimes. We study the
``kink'' mechanism, and we obtain predictions for the wavelength of the
layers as a function of the different parameters of the system. In
particular we find that the wavelength of the layers is proportional to
the flux of adding grains (i.e., proportional to the thickness of the
rolling phase), and we also find a crossover behavior at the thin-thick
flow transition.
Finally we discuss our results in regard to previous approaches to segregation
\cite{degennes} and stratification \cite{mcs} and to the experiments
\cite{makse,makse2,yan,kaka2}. A short report of some of the results
presented here has been published in \cite{mcs}.

\section{Cellular Automaton Model}
\label{discretemodel}

\subsection{Motivation and Definitions}
\label{motiv}

We first introduce a simplified discrete model to understand the
mechanism leading to stratification. Discrete models have been used
before for modeling the complex behavior of granular materials. The
role played by the angle of repose has been used to
understand the complex dynamics of granular flows of mixtures
\cite{varenna,bouchaud,bagnold2,jaeger}. 
The concept of angle of repose normally involves a
macroscopic definition---i.e., the angle of repose is the angle of the
pile surface after the grains are poured onto a heap. Here, we will define a
microscopic angle of repose as the maximum angle at which a rolling
grain is trapped at the surface of the pile. This definition must be
understood in terms of a coarse grained distance along the surface of
the pile---of the order of a few grain sizes---over which all the
quantities involved in the model are averaged.


\begin{figure}[htbp]
\centerline{ \vbox{ \hbox{\epsfxsize=5.cm \epsfbox{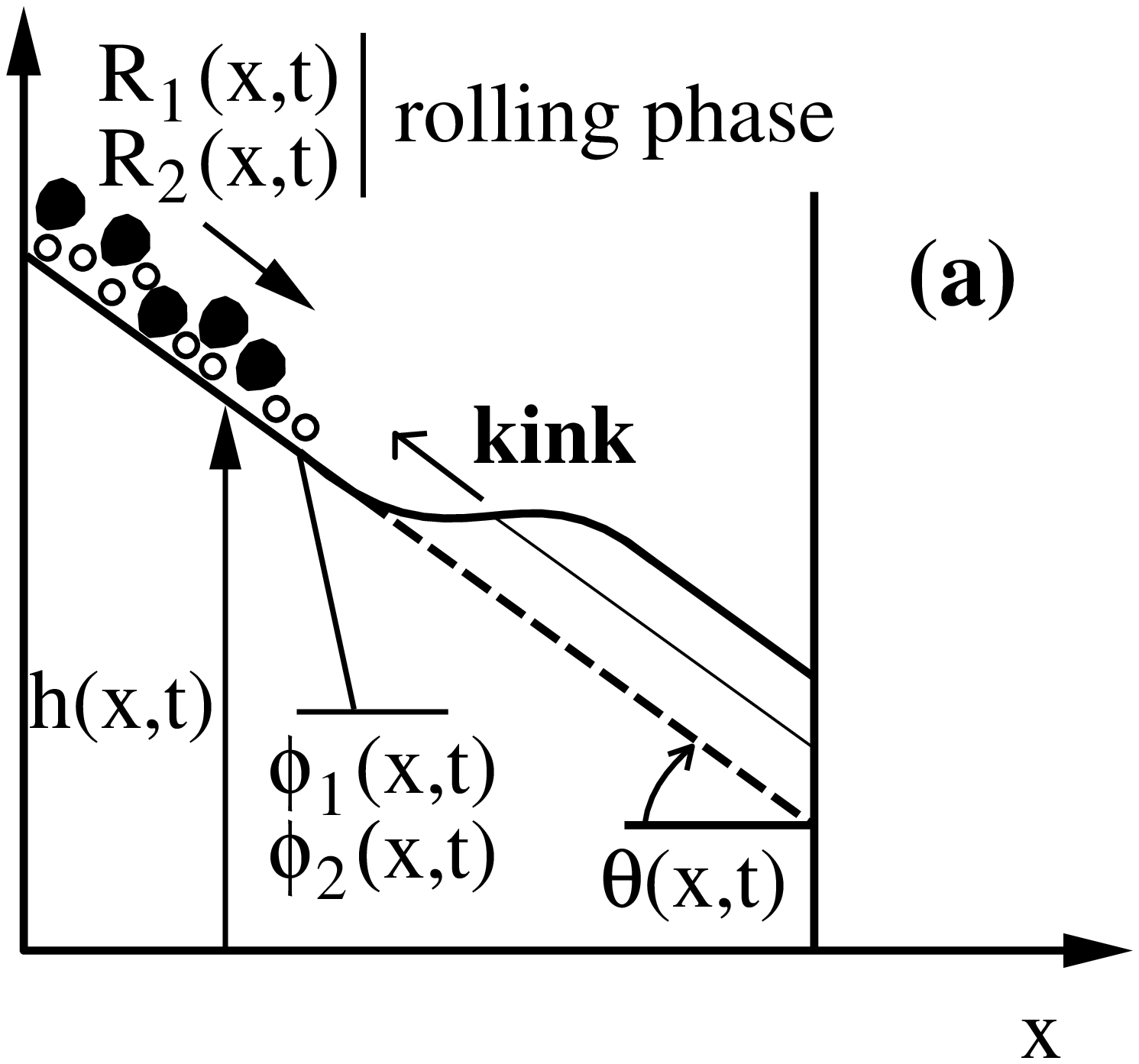} }
 \hbox {\epsfxsize=5.cm \epsfbox{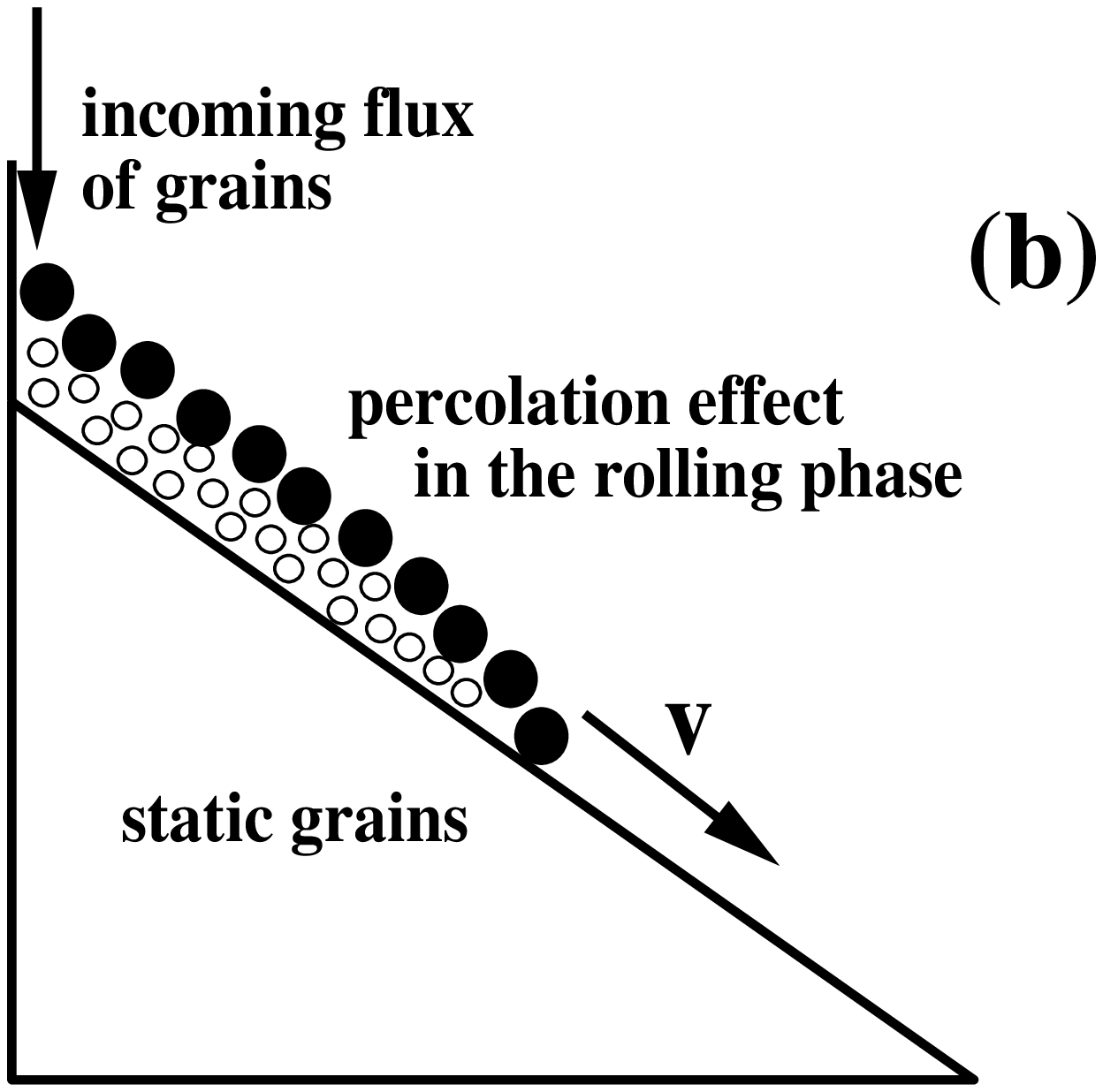} } } 
}
\narrowtext
\caption{(a) 
Typical experimental set-up to study spontaneous stratification and
segregation in quasi-two-dimensional cells, along with the 
different quantities defined in the text.
(b) Schematic representation of the percolation effect in the
rolling phase. We pour continuously a mixture of large and small grain
in a two-dimensional cell and observe a steady thick flow of grains. The
percolation effect consists of the size segregation of grains in the
rolling phase: large grains are observed to rise to the top of the
rolling phase, and small grains drift to the bottom of the rolling
phase. Due to the percolation effect, the small grains are the first to
be captured at the pile surface, resulting in the segregation of the
mixture in the bulk. }
\label{cell}
\end{figure}

In the case of a single-species sandpile, as sand is added to the
sandpile we consider a critical angle of repose $\theta_r$. When the
local angle of the pile is smaller than $\theta_r$ the rolling grain is
stable and there is no flow. When the local angle of the pile is larger
than $\theta_r$ the rolling grain is unstable and flow occurs.

We develop a simple discrete model, based on the idea that grains flow
when the local angle of the sandpile is larger than the angle of repose.
We consider that the angle of repose depends on the type of rolling
grains and also on the composition of the surface. Thus, in the case of
granular mixtures of two different species, we consider the existence of
four different {\it generalized angles of repose.}

The sandpile is built on a lattice, where the grains have the same
horizontal and vertical size as the lattice spacing.
The two species will be distinguished according to different generalized
angles of repose.
In general, we will call species type $1$ to the small grains, and
species type $2$ to the large grains. The species can also have 
 rough or smooth surface 
properties, which,   together with the different  size,
will define the 
different angles of repose of the species.
Following \cite{bouchaud,degennes-french,degennes}, we
regard each grain as belonging to one of two phases:

\begin{itemize}
\item
{\bf Static phase:} if the grain is part of the sandpile,

\item
{\bf Rolling phase:} if the grain is not part of the sandpile and rolls
downward with a constant drift velocity.
\end{itemize}

We consider the local slope
\begin{equation}
s_i \equiv h_{i} - h_{i+1}
\end{equation}
of the static grains as the variable controlling the dynamics of the
rolling grains. Here, $h_i$ denotes the height of the sandpile at
coordinate $i$, and we assume the pouring point at $i=1$.

\begin{figure}[htbp]
\centerline{ \vbox{ \hbox{\epsfxsize=6cm
\epsfbox{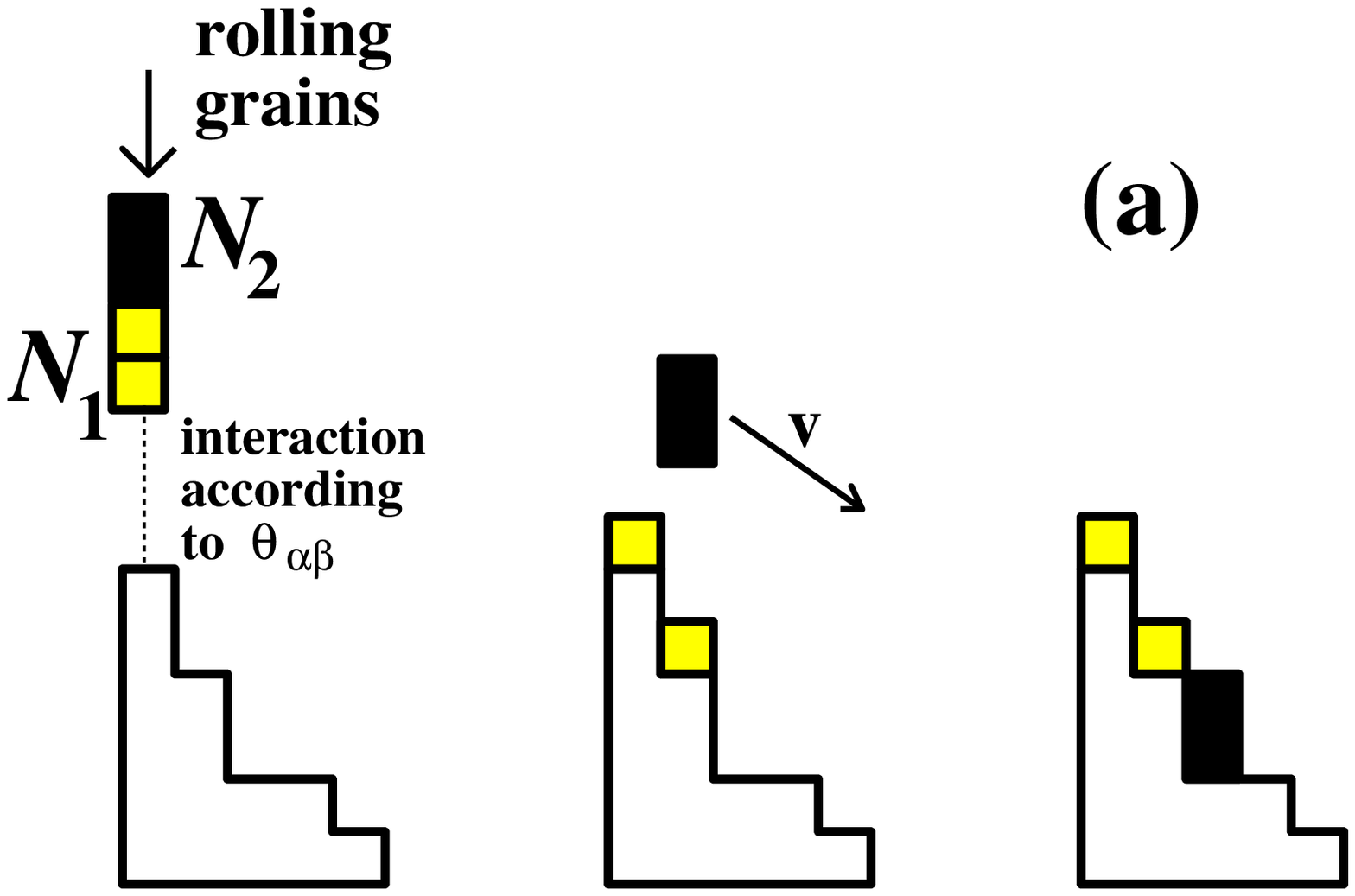} }}}
\vspace{.8cm}
\centerline{ \vbox{ \hbox{\epsfxsize=6.cm \epsfbox{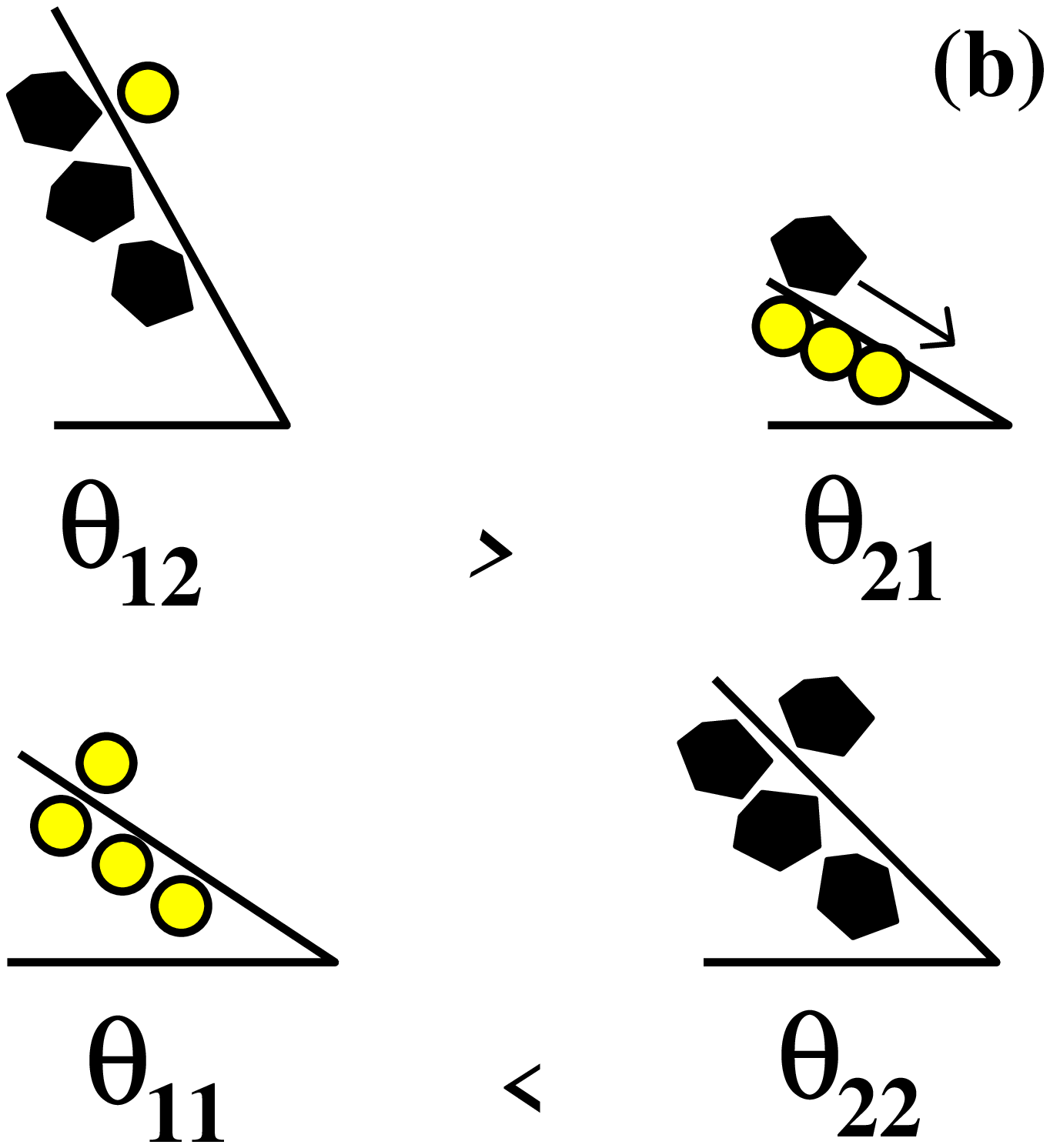} }}
}
\vspace{1cm}
\narrowtext
\caption{(a) Description of the discrete model.
(b) The four dynamical angles of repose $\theta_{\alpha\beta}$ depends
on the composition of grains at the surface of the pile, and are chosen
according of the four possible interactions between the two species
of grains. As discussed in the text, the angles of repose satisfy
$\theta_{21} < \theta_{11} < \theta_{22}< \theta_{12}$, when species $1$
are small and rounded and species $2$ are large and rough, and
$\theta_{21} < \theta_{22} < \theta_{11}< \theta_{12}$, when species $1$
are small and rough and species $2$ are large and rounded.}
\label{descr-micro}
\end{figure}

At each time step a set of $N_1$ small grains and $N_2$ large grains is
deposited at the top of column $1$ of the pile [see
Fig.~\ref{descr-micro}(a)]. These grains are considered to belong to the
rolling phase. In the case of thin flow we assume that the rolling phase
is homogeneous and both types of species are mixed in the fluid phase
and interact with the surface. Then one rolling grain of each species
interacts with the surface at each time step, and can be converted from
the rolling phase to the static phase. In the thick flow regime, the
percolation effect is expected to take place, and the small grains
screen the interaction of the large rolling grains with the pile
surface. Thus in the thick flow regime, we consider that the large
grains interact with the pile surface 
only when the number of small rolling grains
at a given position falls below a given threshold $\epsilon \ll N_2$.

The remaining rolling grains which do not interact with the pile surface
are convected ``downward'' with unit ``drift velocity''---i.e., they
move to the next column at each time step.

The dynamics of each rolling grain interacting with the sandpile surface
is governed by its local angle of repose.
We consider that the repose angle
depends on the local composition on the surface, so we define
$\theta_{\alpha\beta}$ as the generalized 
repose angle of a rolling grain of type
$\alpha$ on a surface with local grains of type $\beta$. We propose that 
\begin{equation}
\theta_{21}<\theta_{12}
\label{t1}
\end{equation}
to take into account the fact that large rolling 
grains roll more easily on top
of small static 
grains than small grains roll on top of large static grains [since the
surface ``looks'' smoother for large grains rolling on top of small
grains, see Fig.~ \ref{descr-micro}(b)]. The repose angles of pure species
$\theta_{\alpha\alpha}$ lie between $\theta_{21}$ and $\theta_{12}$.

Since a large grain roll easier than 
a small rolling grains 
on top of a layer of small static grains, we have 
\begin{equation}
\theta_{21} < \theta_{11}.
\label{t2}
\end{equation} 
Conversely, a large grain roll easier than a small grain on top of a 
layer of large static grains, so that 
\begin{equation}
\theta_{22} < \theta_{12}.
\label{t3}
\end{equation}

The stratification experiments \cite{makse} use a mixture 
smaller less faceted grains and larger more
faceted grains. Thus, the last inequality between the angles of repose is 
provided by the fact that the grains differ in shape of friction properties.
Since the small species are the most rounded, 
then the repose angle of the smaller pure species is 
smaller than the repose angle of the large pure species---i.e.,
\begin{equation}
\theta_{11} < \theta_{22}.
\label{t4}
\end{equation} 
Therefore, 
to mimic the experimental conditions for stratification \cite{makse}, we
propose
\begin{equation}
\theta_{21} < \theta_{11} < \theta_{22}<\theta_{12}.
\label{t5}
\end{equation}
We notice that the conditions (\ref{t1})-(\ref{t3}) are a
consequence of the different size of the species, while the condition
$\theta_{11} \neq \theta_{22}$ is achieved for mixtures of grains
with different shapes, and does not depend on the size of the grains.
Thus the model incorporates the size segregation and shape 
segregation in the definition of the angles of repose. In general the grains
with the larger angle of repose will tend to be captured first. Thus the small
grains will tend to be captured at the top of the pile, and the
smoothest grains at the bottom of the pile. The percolation effects
tend to segregate the small grains at the top of the pile, thus, it acts in the
same way as the size segregation due to different angles of repose
Eqs.  (\ref{t1})-(\ref{t3}).

At each time step, the rolling grain interacting with the sandpile
surface at coordinate $i$ either will stop (by being converted into a
static grain) if the local slope of the surface
\begin{equation}
h_{i} - h_{i+1} \le s_{\alpha\beta}\equiv \tan \theta_{\alpha\beta},
\end{equation}
where $s_{\alpha\beta}$ is the local slope, or will continue to roll
(together with the remaining rolling grains) to column $i+1$ if
\begin{equation}
h_{i} - h_{i+1} > s_{\alpha \beta}.
\end{equation}
We iterate this algorithm to form a large sandpile of typically $10^5$
grains. We assume that the substrate is made of a layer of large
grains.

The discrete model in the thick flow and thin flow regime gives rise to
very similar results. Then in the following we will concentrate only on
the thick flow regime where most of the experiments are done, and where
the grains are segregated in the rolling phase. Indeed, we will see
that the differences between the thin flow regime and the thick flow
regime arise only when cross-amplification processes are taken into
account. Since the discrete model neglects these type of processes, then
it gives rise to similar results for both flow regimes.

\begin{figure}[htbp]
\centerline{ \vbox{ \hbox{\epsfxsize=7cm \epsfbox{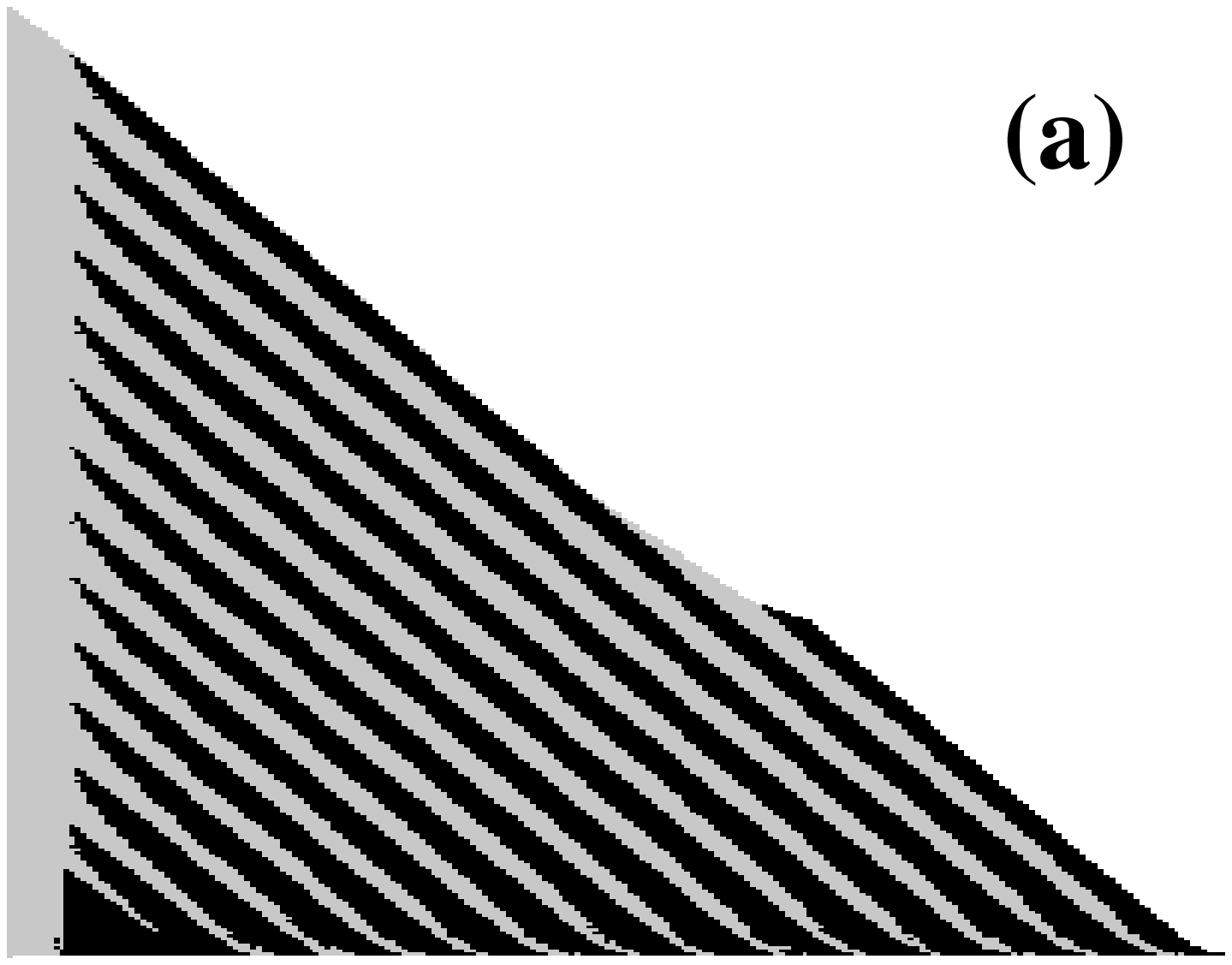}} }
\vspace{0.5cm} } 
\centerline{ \vbox{ \hbox{\epsfxsize=7cm
\epsfbox{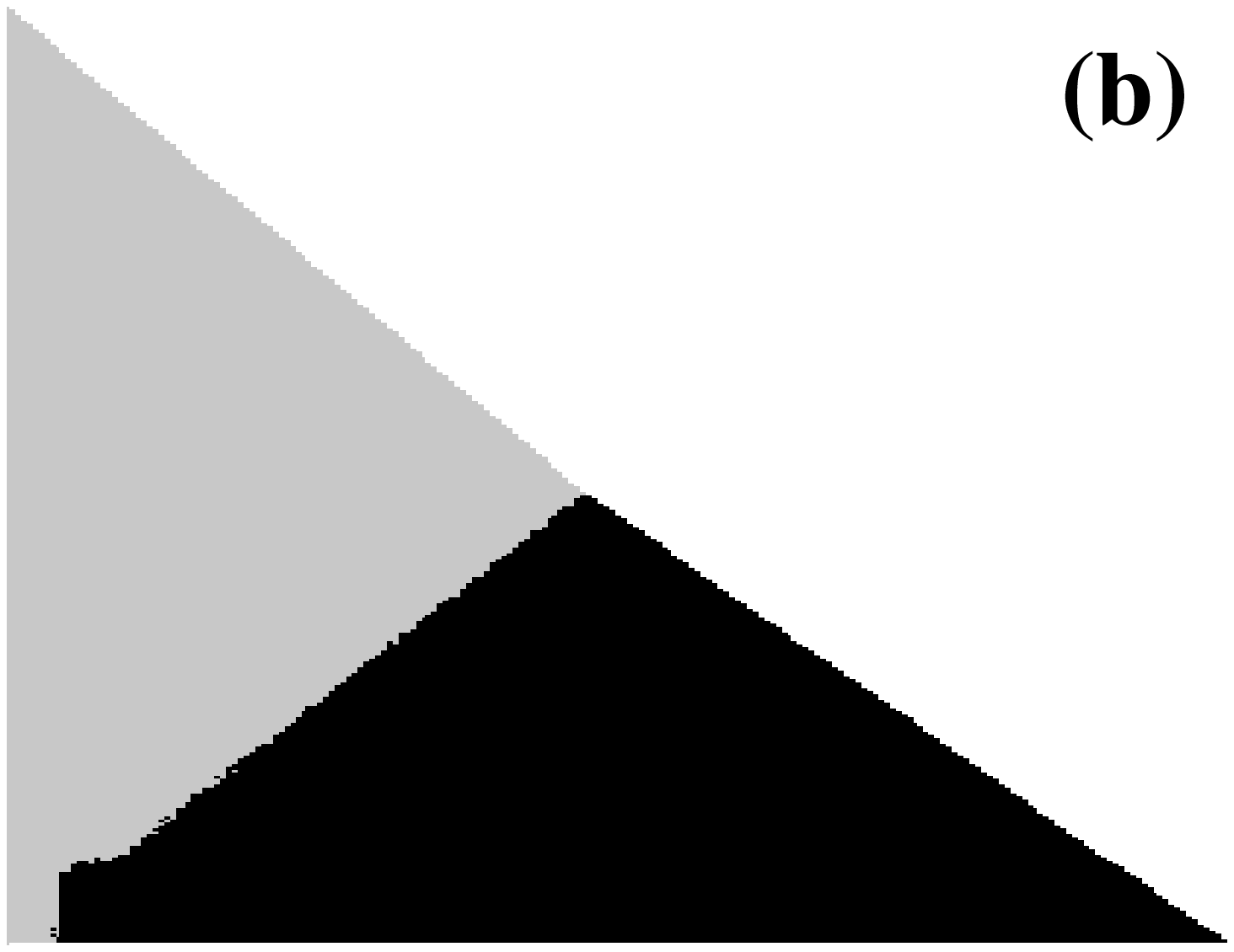}} } 
\vspace{1cm}} 
\narrowtext
\caption{ (a) Spontaneous Stratification: Results obtained with
the discrete model for $N_1=10$, $N_2=10$, $\epsilon = 1$, 
 $s_{11}=4$, $s_{12}=7$,
$s_{21}=2$, $s_{22}=5$. Note
the kink profile at which grains are stopped during an avalanche. The large
rough
grains are black and the small rounded grains are gray. (b) Segregation:
Results obtained with the discrete model when the angle of repose of
the large rounded grains is smaller than the angle of repose of the small
rough 
grains for  $N_1=10$, and $N_2=10$,  $\epsilon = 1$, 
$s_{11}=5$, $s_{12}=7$, $s_{21}=2$,
$s_{22}=4$.}
\label{discrete}
\end{figure}

\subsection{Kink Mechanism}

\begin{figure}
\centerline{ \vbox{ \hbox{\epsfxsize=9cm
\epsfbox{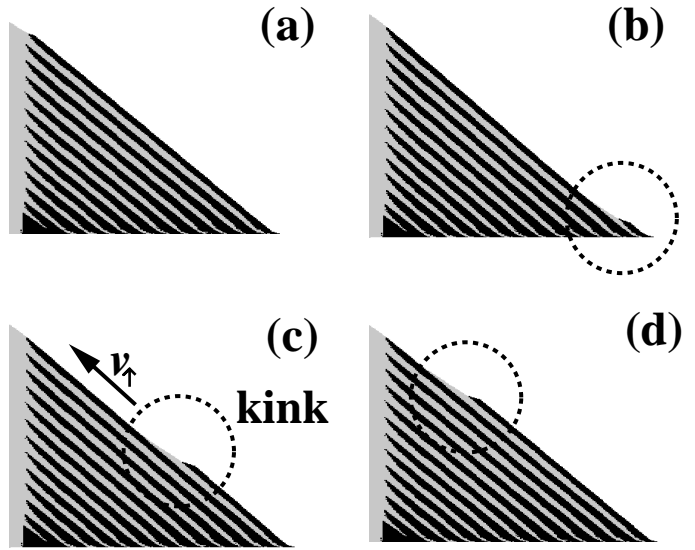}} } \vspace{1cm} } 
\narrowtext
\caption{Dynamics obtained with the discrete model, showing the
formation of a pair of layers at the kink. Notice the kink moving
uphill at constant velocity $v_\uparrow$.}
\label{seq-her}
\end{figure}

Figure \ref{discrete}a shows the resulting morphology predicted by the
discrete model when the large grains are rougher than the small grains.
This fact is quantified by taking into account that the angle of repose
of the pure large species is larger than the angle of repose of the pure
small species, $\theta_{22}>\theta_{11}$. The stratification is
qualitatively the same as that found experimentally \cite{makse}, both
in regard to the {\it statics} of the sandpile [seen in
Fig.~\ref{discrete}(a)], and also in regard to the {\it dynamics\/} (seen
in Fig.~\ref{seq-her}).

The mechanism for spontaneous stratification involves the formation of a
kink at which the grains are stopped during an avalanche. After a pair
of static layers is formed with a layer of large grains on top of a
layer of small grains, the angle of the sandpile is close to
$\theta_{22}$ [Fig. \ref{seq-her}(a)]. 
Since the surface of the sandpile is made of large
grains and $\theta_{22}<\theta_{12}$, a thin layer of small grains is
trapped on top of the layer of large grains as more grains are poured
onto the cell. These small grains smooth
the surface without changing significantly the sandpile angle, and allow
rolling small grains to go further down (since
$\theta_{11}<\theta_{22}$)  [Fig. \ref{seq-her}(b)]. 
 The large grains are rolling down on this
thin layer of small grains without being captured
($\theta_{21}<\theta_{22}$), and are the first to reach the bottom of
the sandpile, giving rise to segregation  [Fig. \ref{seq-her}(b)]. 
When the flow reaches the
base of the pile, the grains develop a profile which displays a kink
where the grains are stopped: the small grains are stopped first at the kink
since
$\theta_{21} < \theta_{11}$, and a pair of layers begins to form, with
the small grains underneath the large grains  [Fig. \ref{seq-her}(c)]. 
 The kink moves upward (in
the opposite direction to the flow of grains) until it reaches the top
of the pile and a complete pair of layers has been formed
 [Fig. \ref{seq-her}(d)].

If, on the other hand, we consider $\theta_{22} < \theta_{11}$ in the
model---corresponding to a mixture of smaller ``more faceted'' grains,
and larger ``less faceted'' grains---we find the complete segregation of
the mixture but no stratification [Fig.~\ref{discrete}(b)]. In this
case, the small and faceted grains are to be found near the top of the
pile, while the large and rounded grains are found near the bottom of
the pile. Thus, the control parameter for spontaneous stratification
appears to be the difference in the repose angle of the pure species, which 
quantifies the difference in shape of the grains.

As seen in Fig.~\ref{discrete}(a), before the layers appear there is an
initial regime where only segregation is found. At the onset of the
instability leading to stratification, a few large grains are captured
on top of the region of small grains near the center of the pile where
the angle of the pile is $\theta \simeq \theta_{11}$. The repose angle
for large rolling 
grains is now $\theta_{22}$. Thus, if $\theta \simeq
\theta_{11} < \theta_{22}$, more large grains can be trapped (since
the angle of the surface is smaller than the repose angle), leading to
the first sub-layer of large grains and then to stratification. On
the other hand, if $\theta \simeq \theta_{11} > \theta_{22}$, no more
large grains can get
captured, the fluctuation disappears, and the
segregation profile remains stable. Thus this picture suggests that,
in agreement with \cite{makse}, the segregation profile observed in
the initial regime is ``stable'' so long as $\theta_{22}<
\theta_{11}$, and unstable (evolving to stratification for large
enough systems) when $\theta_{11} < \theta_{22}$. This explains
qualitatively why the control parameter for the
stratification-segregation transition is the angle of repose of the
pure species.

\section{Continuum Theory for granular flow}
\label{continuum}

A theoretical approach to segregation in granular flow is not a simple
task. On one hand, extensive numerical simulations (such as molecular
dynamics \cite{bideau,wolf,cundall,grest,ristow,baumann}, or lattice gas models
\cite{antal1,hans1,hans2,antal2}) have been able to reproduce
several segregation phenomena observed experimentally. On the other
hand, an analytical approach of segregation, considering granular medium
as continuum, would be instructive, since it would allow to reduce the
problem to a small set of parameters that control the system, and then
to get a clear phenomenological understanding of the problem. However,
an important question is whether the grains can be treated as a
continuum medium. A continuum approach means that we will be able to
replace the grains belonging to a same region of space by some average
quantities, for example the density, or the averaged speed. But specific
grains can play a peculiar role, i.e., in the case of arching effects,
where an arrangement of several grains in a special way (an arch)
supports the above grains, prevents the flow by gravity, and creates
large discontinuities for the local density or the force
\cite{arching,wittmer}.

In this section, we study spontaneous stratification using a continuum
formalism. We will show that the results obtained are very close to the
experiment, and then we extract the important parameters of the problem.
Our results shows that a continuum formalism is able to reproduce a
complex phenomenon in granular matter such as spontaneous
stratification.

The continuum theory of granular flow for a single-species sandpile was
proposed by Mehta {\it et al.} \cite{metha} and Bouchaud {\it et al.}
\cite{bouchaud}. Bouchaud {\it et al.} recognized the necessity of
treating differently the rolling grains (rolling on the surface of the
surface), and the static grains that form the sandpile. The rolling
grains are in a ``liquid'' state, flowing down by gravity and
interacting with the static grains in the ``solid'' state. The dynamics
of the sandpile is then given by how the rolling grains move on the
surface and how they interact with the sandpile. The equations for those
two quantities for the case of single-species sandpiles are
 
\begin{mathletters}
\begin{eqnarray}
\label{lagrange_R}
\frac{d}{dt}R[x(t)] & = & \Gamma(R,h),\\ \frac{\partial
h(x,t)}{\partial t} &=& -\Gamma(R,h).
\label{lagrange_h}
\end{eqnarray}
\label{lagrange}
\end{mathletters}
Here $R[x(t)]$ is the height of the layer of rolling grains in the
Lagrange representation (following the rolling grains), and $h(x,t)$ the
height of the sandpile [Fig.~\ref{cell}(a)]. The rolling grains
interact with the sandpile through $\Gamma(R,h)$, which represents the
height of rolling grains becoming part of the sandpile. Equations
(\ref{lagrange}) conserve the number of grains.

We need to define how the rolling grains move on the surface, and
how they interact with the sandpile. Two forces are acting on the
rolling grain: (i) gravity and (ii) friction coming from the collisions
of the rolling grain with the grains of the sandpile. The number of
collisions at the scale of interest for continuum equations (several
times the size of a grain) is large enough to say that at this scale,
the speed is always at its limit value where the two forces balance
exactly, and the motion of a grain is overdamped. Equations
(\ref{lagrange_R}) and (\ref{lagrange_h}) can be rewritten in Euler
representation (looking at the height of grains at a given position $x$)
\begin{mathletters}
\begin{eqnarray}
\frac{\partial R(x,t) }{\partial t} & = & - v \frac{\partial R(x,t)}
{\partial x}
 + \Gamma(R,h), \\
\label{eq:bouchaud_R}
\frac{\partial h(x,t)}{\partial t} & = & -\Gamma(R,h).
\label{eq:bouchaud_h}
\end{eqnarray}
\label{eq:bouchaud}
\end{mathletters}
Here $R(x,t)$ is now a function of the time and the position $x$, and we
assume the pouring point at $x=0$. The rolling grains are moving down
at a speed $v(x,t)$ in the positive $x$ direction, where $v(x,t)$ could
depend on the local slope of the sandpile. However, in the spirit of
capturing only the essential physical mechanisms to recover the observed
phenomena, we will consider the speed $v$ to be constant all over the
sandpile.

Bouchaud {\it et al.} \cite{bouchaud}
proposed a form for $\Gamma(R,h)$ in the general
case, that has been simplified by de Gennes \cite{degennes-french} for
the simpler case of a continuous flow of rolling grains. In this case,
there is no discrete avalanching and only the repose angle must be
included in the formalism \cite{bouchaud.vs.degennes}. The interaction
term $\Gamma(R,h)$ is
\begin{equation}
\label{eq:gamma}
\Gamma(R,h) = \gamma [ \theta(x,t)-\theta_r ] R,
\end{equation}
where
\begin{equation}
\theta(x,t)\equiv \frac{\partial h(x,t)}{\partial x}
\end{equation}
 is the local angle of the surface (we will make no distinction
between the angle of the surface and the tangent of the angle).

Equation (\ref{eq:gamma}) states that the rate of the interaction is
proportional to the number of grains interacting with the sandpile.
Rolling grains will become part of the sandpile if the angle of the
surface $\theta(x,t)$ is smaller than the repose angle $\theta_r$
(``capture''), while static grains will become rolling grains if
$\theta(x,t)$ is larger than $\theta_r$ (``amplification''). The
constant $\gamma$ is the inverse of the time on which this interaction
is effective because the distance on which a rolling grain is stopped
when $\theta(x,t)$ is smaller than $\theta_r$ is $v/\gamma$. This
distance must scale with the size of the grain $d$.  Thus
\cite{degennes-french}
\begin{equation}
 v/\gamma \sim d.
\label{voverg} 
\end{equation} 

The linear dependence of $\Gamma(R,h)$ on the quantity
$[\theta(x,t)-\theta_r]$ can be understood as a first order development
of a more complicated function of $\theta(x,t)$. As soon as the angle of
the sandpile is far from the repose angle, non-linear terms must be
added (avoiding the non-physical divergences that could be found with
the linear development). Finally the proportionality of $\Gamma(R,h)$ to
$R(x,t)$ is meaningful in the case where all the grains interact at each
time with the surface (thin-flow limit). This imposes the height of the
rolling layer to be of the order of the grain size $R(x,t) \simeq d.$
However we will argue that this approximation might be still valid in
the case of thick flows as well, since the interaction might be
proportional to the pressure exerted by the fluid phase, which in turn
is proportional to $R(x,t)$ for a fluid
\cite{zik}.

To treat the problem of segregation in granular flow composed of binary
mixtures, Boutreux and de Gennes (BdG) \cite{degennes} have extended
this formalism to the case of flows made of two types of grains,
identifying three variables: the two heights of rolling grains $R_\alpha
(x,t)$
 (i.e. the total  thickness of the rolling phase
 multiplied by the
local volume fraction of the $\alpha$ grains in the 
rolling phase at position $x$),
and the height of the sandpile $h(x,t)$. BdG generalized Eqs.
(\ref{eq:bouchaud}) to
\begin{mathletters}
\begin{eqnarray}
\frac{\partial R_\alpha}{\partial t} & = & - v_\alpha \frac{\partial
R_\alpha}{\partial x} + \Gamma_\alpha, \mbox{~~~~~~~ $\alpha=1,2$} \\
\frac{\partial h}{\partial t} & = & - \Gamma_1 - \Gamma_2,
\end{eqnarray}
\label{eq:R_et_h}
\end{mathletters}
where the interaction term $\Gamma_\alpha$ takes into account the
conversion of the rolling grains into static grains, and is defined
through the collision matrix $M_{\alpha \beta}$

\begin{equation} 
\Gamma_\alpha \equiv \sum_{\beta=1}^2 M_{\alpha \beta}R_\beta .
\end{equation}

The collision matrix, governing capture and amplification, depends on
the local angle of the pile $\theta(x,t)$ and on the local composition
of the surface of the sandpile $\phi_\alpha(x,h)$, which is a function
of $x$ and $h$. However, writing an equation of evolution for
$\phi_\alpha(x,h)$ is not easy. When both rolling species are
captured, the height of the sandpile increases, and $\phi_\alpha(x,h)$
is given by
\begin{equation}
\label{eq:phi}
\phi_\alpha(x,h) = -\frac{\Gamma_\alpha}{\partial h / \partial t},
\end{equation}
and
\begin{equation}
\phi_1+\phi_2=1.
\end{equation}
As soon as amplification dominates for one or both species,
Eq.~(\ref{eq:phi}) is no longer valid. When $\Gamma_\alpha>0$ for both
species, the height of the sandpile decreases, and $\phi_\alpha(x,h)$
does not have to be updated. Finally, when grains of type $\alpha$ are
captured and grains of type $\beta$ are amplified, the composition will
be $\phi_\alpha(x,h)=1$.

The general form of the collision matrix is defined by taking into
account a set of binary collisions between a rolling and a static
grains \cite{degennes}

\begin{equation}
\label{canonical}
\hat{\mbox{M}} \equiv \left (
\begin{array}{lr}
a_1(\theta) \phi_1 - b_1(\theta) & x_2(\theta) \phi_1 \\ x_1(\theta)
\phi_2 & a_2(\theta) \phi_2 - b_2(\theta)
\end{array}
\right ) .
\end{equation}
This definition involves a set of a priori unknown collision functions
contributing to the rate processes: $a_\alpha(\theta)$ is the
contribution due to an amplification process (when a static grain of
type $\alpha$ is converted into a rolling grain due to a collision by
a rolling grain of type $\alpha$), $b_\alpha(\theta)$ is the
contribution due to capture of a rolling grain of type $\alpha$ (when
a rolling grain of type $\alpha$ is converted into a static grain),
and $x_\alpha(\theta)$ is the contribution due to a
cross-amplification process, (the amplification of a static grain of
type $\beta$ due to a collision by a rolling grain of type $\alpha$).

BdG \cite{degennes} proposed an analytically tractable form for the
collision matrix which includes capture and amplification, in the case
where the two species have the same size but differ in respect to their
repose angle. BdG found the steady state solution in the geometry of
the silo, in the case where the repose angle does not depend on the
composition of the surface. The BdG model explains the segregation of
the two type of grains in different regions of the sandpile, but not the
phenomenon of spontaneous stratification. A generalization of the model
to include also the different sizes of the grains \cite{thomas} shows a
smooth segregation of the species with the concentrations of static
grains behaving as a power law of the position along the surface pile;
results valid only when the species do not differ too much (a fact that
allows to perform linear approximations of the collision functions
around the angles of repose \cite{thomas}). In the following we propose
a form of the collision functions valid when there is a large difference
in size and shape between the species and we treat the thin and thick
flow regimes, in order to understand stratification and segregation as
seen in the experiments.

\begin{figure}[htbp]
\centerline { \vbox{ \hbox {\epsfxsize=8cm \epsfbox{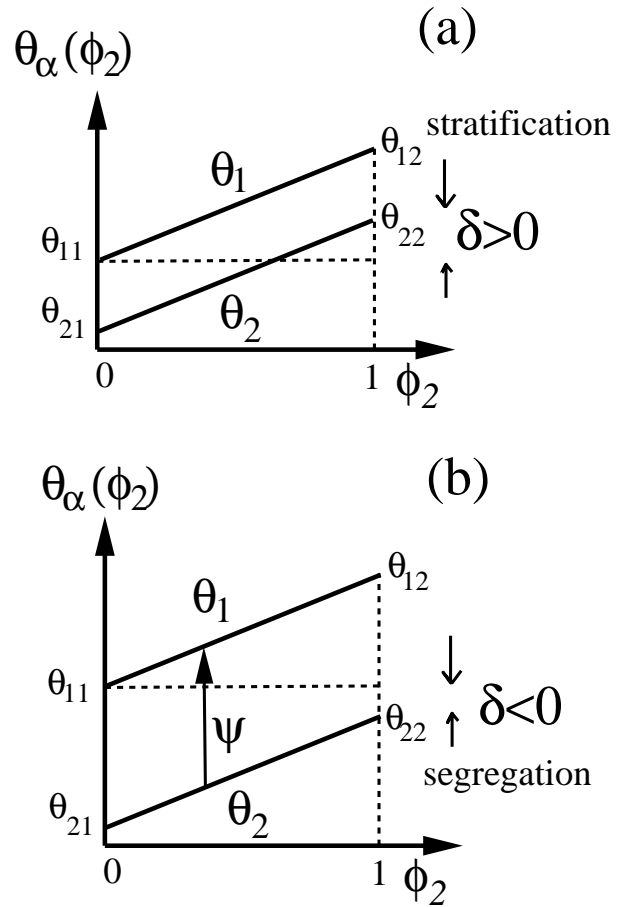} }
} } 
\vspace{1cm}
\narrowtext
\caption{Dependence of the repose angle for the two types of rolling
grains on the concentration of the surface of large grains
$\phi_2$. (a) An essential ingredient to obtain segregation is that
$\theta_{22} > \theta_{11}$, (b) while for segregation we require
$\theta_{11} > \theta_{22}$. For the numerical integration, we use the
linear interpolation between $\phi_2=0$ and $\phi_2=1$ as plotted
here.}
\label{contia}
\end{figure}

\section{Continuum Model for Stratification of Granular Mixtures}

\subsection{The collision matrix and the generalized angle of repose}
\label{collision}

We present the collision matrix $M_{\alpha \beta}$ that includes the
effects of capture and amplification, and the dependence of the repose
angle on the composition of the surface as discussed in \cite{mcs}.
Considering one type of rolling grain $\alpha$, we define capture and
amplification as follows: if the local angle of the surface
$\theta(x,t)=-\partial h(x,t) /\partial x$ is smaller than
the generalized angle of repose of the rolling specie $\alpha$
$\theta_\alpha$, the $\alpha$ rolling grains will be captured. In the
reverse case $\theta(x,t)> \theta_\alpha$, the static grains at position
$x$ will be amplified. Amplification will cause locally a small hole,
i.e., the two types of static grains will be amplified the same way
according to their local concentration $\phi_\alpha$.

The generalized repose angle $\theta_\alpha$ of each type of rolling
grain is now a continuous function of the composition of the surface
$\theta_\alpha=\theta_\alpha (\phi_\beta)$ (see Fig.~\ref{contia})

\begin{equation}
\label{dependence}
\begin{array}{rcl}
\theta_1(\phi_2) &=& m \phi_2 + \theta_{11}\\ \theta_2(\phi_2) &=& m
\phi_2 + \theta_{21} = - m \phi_1 + \theta_{22},
\end{array}
\end{equation}
where $m\equiv \theta_{12} - \theta_{11} = \theta_{22} - \theta_{21}$.
We have assumed the difference
\begin{equation}
\psi \equiv \theta_1(\phi_2)-\theta_2(\phi_2)
\end{equation}
to be independent of the concentration $\phi_2$. For simplicity, we
choose a linear interpolation between the extreme values
$\theta_\alpha(\phi_\beta =0)$ and $\theta_\alpha(\phi_\beta =1)$. If we
label the small grains with $\alpha=1$ and the large grains with
$\alpha=2$ and denote the extreme values of the repose angle by
$\theta_\alpha(\phi_\beta=1)=\theta_{\alpha \beta}$ as in
Sec.~\ref{discretemodel}, the difference in size implies [see
Fig.~\ref{descr-micro}(b)] 
\beq 
\theta_{12}<\theta_{21}. 
\eeq

If
species 1 are the smallest then $\theta_1(\phi_2)>\theta_2(\phi_2)$ for
any composition of the surface $\phi_2$--- i.e., the small grains are
always the first to be captured. 
Thus, the angular difference $\psi$ is determined mainly 
by the difference in size
between the species, and the value of 
 $\psi$ determines the degree
of segregation; the larger $\psi$, the stronger the segregation. 
The repose angles of the pure species
$\theta_{11}$ and $\theta_{22}$ will be intermediate between
$\theta_{12}$ and $\theta_{22}$, and their relative values will depend
on the shape of the two species as discussed in Sec. \ref{motiv}.

For mixtures of grains with different shapes or friction coefficients we
have $\theta_{11}\neq\theta_{22}$, and $\theta_{12} = \theta_{21}$ if
the species have the same size. If $\theta_{11}$ is the repose angle of
the pure round species and $\theta_{22}$ is the repose angle of the pure
rough species, then $\theta_{11}<\theta_{22}$. If the species have the
same shape or friction coefficients then $\theta_{11}=\theta_{22}$.

When the size ratio is close to one ($d_2/d_1 \stackrel{<}{\sim} 1.5$),
the angle $\psi$ is expected to be small and then it is plausible to
linearize the collision functions around the angles of repose as done in
\cite{thomas}. When the is a larger difference in size ($d_2/d_1
\stackrel{>}{\sim} 1.5$), we expect $\psi$ to be large, and we 
approximate the collision functions and define the following 
collision matrix:
%
\begin{equation}
M_{\alpha \beta}= \left (
\begin{array}{cc}
\gamma_{11}\delta\theta_1 \Theta_{\mbox{\tiny self}} [\delta\theta_1,
\phi_1] &
\begin{array}{lr}
\gamma_{12}\delta\theta_2\Theta_{\mbox{\tiny cross}} [\delta\theta_2,
\phi_1]
\end{array}
\\ \\ \gamma_{21}\delta\theta_1\Theta_{\mbox{\tiny cross}}
[\delta\theta_1, \phi_2] &
\gamma_{22}\delta\theta_2\Theta_{\mbox{\tiny self}} [\delta\theta_2,
\phi_2]
\end{array}
\right ),
\label{canonical1}
\end{equation}
where
\begin{equation}
\delta \theta_\alpha \equiv \theta(x,t) - \theta_\alpha.
\end{equation}
Here, $\gamma_{\alpha \alpha} > 0$ has dimension of inverse time, and $v
/ \gamma_{\alpha \alpha}$ represents the length scale at which a rolling
grain will interact significantly with a surface at an angle above or
below the angle of repose.

The functions $\Theta(\delta \theta_\alpha, \phi_\alpha)$ distinguish
capture from amplification. $\Theta_{\mbox{\tiny self}}$ treats the
case of self amplification, i.e., amplification of static grains
$\alpha$ by rolling grains $\alpha$, and $\Theta_{\mbox{\tiny cross}}$
the case of cross-amplification, i.e., amplification of static grains
$\alpha$ by rolling grains $\beta$. More specifically, we consider
\beq
\Theta_{\mbox{\tiny self}} (\theta-\theta_\al,
\phi_\al)= \left \{
\begin{array}{ll}
1 & {\ds ~~ \mbox{ if ~ $\theta -\theta_\alpha < 0 $} } \\ \phi_\alpha &
{\ds ~~ \mbox{ if ~ $\theta -\theta_\alpha > 0 $} }
\end{array}
\label{Theta}
\right . .
\end{equation}
These equations mean that in the case of capture, the amount of rolling
grains of type $\alpha$ converted to static grains is proportional to
the number of grains interacting with the surface, i.e., $R_\alpha$; in
the case of amplification, the amount of static grains converted to
rolling grains is proportional to $R_\alpha$ again and to the number of
static grains $\alpha$ on the surface, i.e., $\phi_\alpha$. Accordingly,
we obtain \beq \Theta_{\mbox{\tiny cross}} (\theta-\theta_\al,
\phi_\beta)= \left \{
\begin{array}{ll}
0 & {\ds ~~ \mbox{ if ~ $\theta -\theta_\alpha < 0 $} } \\ \phi_\beta
& {\ds ~~ \mbox{ if ~ $\theta -\theta_\alpha > 0 $} }
\end{array}
\right . .
\end{equation}

In terms of the collision functions defined in (\ref{canonical}) we
have
\begin{mathletters}
\label{canonical2}
\begin{equation}
\begin{array}{llcl}
a_\alpha(\theta)&\equiv& \gamma_{\alpha\alpha} & \Pi[\theta(x,t)-
\theta_\alpha(\phi_\beta)], \\ b_\alpha(\theta)&\equiv&
\gamma_{\alpha\alpha} & \Pi[\theta_\alpha(\phi_\beta)-\theta(x,t)], \\
x_\beta(\theta) &\equiv& \gamma_{\beta\alpha} & \Pi[\theta(x,t)
-\theta_\beta(\phi_\beta)],
\end{array}
\end{equation}
where
\begin{equation}
\Pi[x] \equiv \left \{
\begin{array}{cr}
0 &~~~~~~~\mbox{if $x < 0$} \\ x &~~~~~~~\mbox{if $x \ge 0$}
\end{array}
\right . .
\end{equation}
\end{mathletters}

This form for the matrix $M_{\alpha \beta}$ supposes that amplification
(respectively capture) occurs only when the local angle is larger
(respectively smaller) than the repose angle. Thus, for a given angle
only capture or amplification can occur at a given time;
an assumption which is expected to be valid when the  grains differ 
appreciable ($\psi$ large).
This fact gives
rise to strong segregation effects with exponential behavior
of the concentrations
as we will corroborate when calculating the steady state solution
of the problem in Sec. \ref{steady}.
In contrast, the form of the
collision functions proposed in \cite{degennes,thomas} assumes that
amplification and capture are linear functions of the local angle. For  
a given angle both amplification and capture act simultaneously, and
the
repose angle corresponds to the angle where amplification and capture
are equal. This linear approximation is valid when the species do not
differ much (for small $\psi$), and gives rise to a weaker segregation
of the species--- with a power law behavior of the concentrations as a
function of the position along the pile surface--- than the segregation
found in this study which is valid when the species differ appreciable
in size and shape.

We focus here on the dependence of the repose angles on the composition
of the surface. We will consider that the other parameters---i.e.,
$\gamma_{\alpha\alpha}$ and $v_\alpha$---do not depend on the
composition of the surface and on the species considered
\begin{equation}
\gamma_{11}=\gamma_{22}=\gamma, ~~~~~v_1=v_2=v.
\end{equation}

In the conditions of the experiment \cite{makse}, where an equal volume
of the two species is poured at the left side of the cell, the boundary
conditions are
\begin{equation}
R_\alpha(0,t)=R_\alpha^0= \frac{R^0}{2}.
\end{equation}

These equations are meaningful if all the rolling grains interact each
time with the surface, and do not interact directly with each other.
Then, when there is percolation in the rolling phase, the interaction
terms have to be modified accordingly. In the next section we will
show how to introduce the percolation effect characteristic of the
thick flow regime. However, we will show that the above equations---
although strictly valid in the thin-flow limit, i.e., $R_\alpha \simeq
d_\alpha$--- will be also needed to describe completely the thick flow
regime.

\subsection{Thick flow regime and the percolation effect }
\label{percolation}

An important consideration of the model proposed above is that the two
types of rolling grains do not interact significantly each other when the
flow of grains is small, i.e., $R_\alpha(x,t) < \epsilon$, where
$\epsilon$ is a cut-off of the order of the grain size $d_\alpha$. This
hypothesis may not be valid in experiments where the input flux is large,
where the layer of rolling grains could be larger than one grain size.

In the case of a thick layer of rolling grains, the phenomenon of
percolation occurs; during the flow of grains down an inclined plane,
the small rolling grains percolate downward through the gaps left by the
motion of larger rolling grains \cite{drahun,savage,makse2}
[Fig.~\ref{cell}(b)]. Therefore, as they are convected down, the rolling
grains segregate, the concentration of small grains being large at the
bottom of the rolling layer, and the concentration of large grains large
at the top of the rolling layer. As only the lowest grains of the
rolling layer interact with the sandpile, the above effect imposes that
small grains will preferentially interact with the static grains. We
include the percolation effect in the continuum model and see how the
results related to spontaneous stratification are affected.

To model this effect, we consider the extreme situation where the
segregation is present as soon as the grains start to roll, i.e., up to
the top of the sandpile, and that only the small grains will interact
with the sandpile as soon as $R_1(x,t)> \epsilon$, with $\epsilon \ll
R^0$. $\epsilon$ is thus the height of small grains below which the
percolation stops to be effective, and for which the large rolling
grains start to interact with the surface. The interaction terms discussed
in Sec. \ref{collision} have
to be modified accordingly.

\begin{figure}[htbp]
\centerline { \vbox{ \hbox {\epsfxsize=7cm \epsfbox{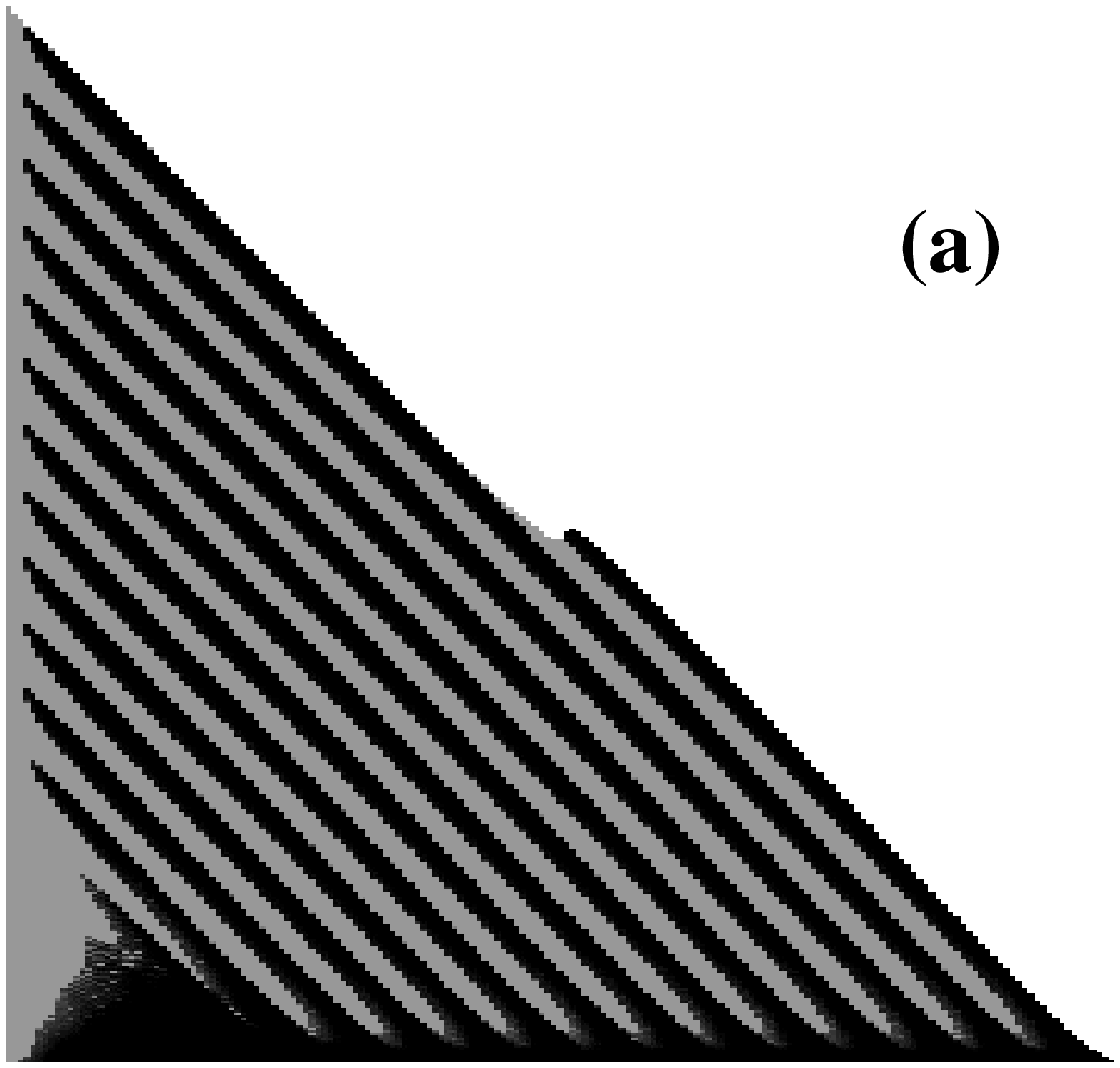}
}}}
\centerline { \vbox{ \hbox {\epsfxsize=7cm \epsfbox{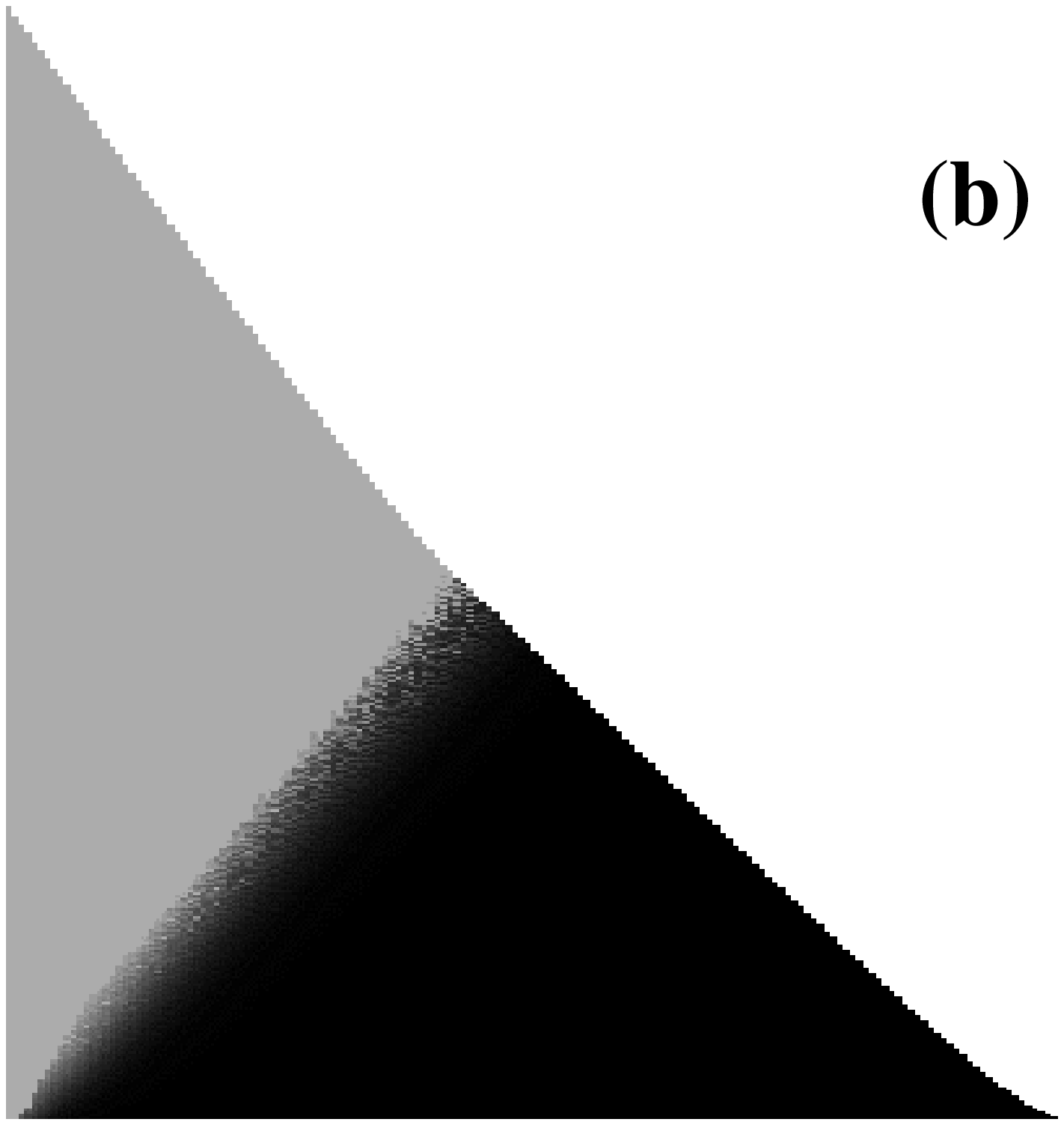}
}}}
\vspace{1cm}
 \narrowtext
\caption{ Thick flow regime: Morphology resulting from the numerical
integration of the continuum equations when the percolation effect is
present. (a) Stratification of a mixture of large rough grains and small
smooth grains. 
The parameters used are $R_1^0=R_2^0=1$, $\epsilon=0.25$, 
$\tan(\theta_{11})=0.5$,
$\tan(\theta_{22})=0.6$, $\tan(\theta_{12})=0.9$,
$\tan(\theta_{21})=0.2$,  $\gamma_{11}$ $=
\gamma_{22}= 1$, $\gamma_{21}$=$\gamma_{12}=0.1$,  and
$v_1=v_2=1$. 
 The dynamics obtained by the numerical
integration of the equation of motion proposed in the text show
the formation of a pair of layers through the kink mechanism.
We also see that the rolling grains are stopped at
the kink in similar fashion as was observed in the stratification
experiment \protect\cite{makse} and in the discrete model in Fig. 
\protect\ref{seq-her}.
(b) Complete segregation of a mixture of small rough grains and large smooth grains
obtained  with the same parameters as in (a) 
except for the angle of repose of the pure species  $\tan(\theta_{11})=0.7$,
$\tan(\theta_{22})=0.5$. 
Notice the total segregation of the mixture. The region
of mixing is concentrated in a small region in the center of the pile. }
\label{thick-results}
\end{figure}

This model implies that except for a small region where
$R_1<\epsilon$, the surface of the sandpile is always divided in two
regions. The upper region corresponds to the part of the surface where
small rolling grains are present. As they are the only one to interact
with the sandpile, the sandpile will be made only of small grains. In
the lower part, only large grains are present (except for the $\epsilon$
part of small grains). The division point between the two regions is
able to move with time.

For the upper part of the sandpile, defined by $R_1(x,t)>\epsilon$, only
small grains are captured, with $R_2(x,t)=R_2^0=R^0/2$ and
$\phi_1(x,t)=1$. The equations become
 
\begin{mathletters}
\begin{eqnarray}
\frac{\partial R_1}{\partial t} & = & -v\frac{\partial R_1}{\partial
x} + \gamma (\theta-\theta_{11})(R_1+R_2) \\ \frac{\partial h}{\partial t} &
= & - \gamma (\theta-\theta_{11})(R_1+R_2)
\end{eqnarray}
\label{eq:percoUpperPart}
\end{mathletters}

When $R_1(x,t)< \epsilon$, the percolation effect disappears and the
equations of evolution of $R_\alpha$ and $h$ can be considered to be the
same as the one defined by the collision matrix (\ref{canonical1}) valid
for the thin flow limit.

The effectiveness
of the interaction (given by $\gamma$ before) must be multiplied now
by the vertical pressure acting on the lowest rolling grain interacting
with the surface, i.e., the weight of the column of rolling grains
above the interacting grain
\cite{zik}. This explains the presence of the $R_2(x,t)$ terms in the
interaction term $\Gamma_1$ in Eq.~(\ref{eq:percoUpperPart}).

We numerically solve Eqs. (\ref{eq:R_et_h}) with the interaction terms
given by (\ref{canonical1}) when $R_1(x,t)<\epsilon$ and by
(\ref{eq:percoUpperPart}) when $R_1(x,t)>\epsilon$ as discussed above.
We find qualitatively the same results as in the experiments
\cite{makse}, with stratification appearing when the large grains are
rougher than the small grains 
[$\theta_{22}>\theta_{11}$, Fig.~\ref{thick-results}(a)]. 
The dynamics of stratification show also
the formation of the kink as observed in the experiments and in the
discrete model shown in Fig.~\ref{seq-her}.
When the
small grains are rougher than the large grains ($\theta_{11}>
\theta_{22}$), then we obtain the strong segregation of the mixture
with the small rough grains found at the top of the heap
(Fig.~\ref{thick-results}(b)) and a small region
of mixing in the center of the pile \cite{time}.

\subsection{Thin flow regime}
\label{thin}

In the absence of the percolation effect, Eq.~(\ref{canonical1}) remains
valid along all the pile since in the small flow limit a thin rolling
phase is expected with $R_\alpha(x,t) < \epsilon$. However, we assume
a large difference in size, so that strong segregation effects are
expected anyway, and the collision functions are expected to behave as
in (\ref{canonical1}). We then solve
Eqs.~(\ref{eq:R_et_h})--(\ref{canonical1}) numerically.
Defining the ``control parameter''
\begin{equation}
\delta \equiv \theta_{22} -\theta_{11},
\end{equation}
we obtain stratification when $\delta>0$, i.e., when the large grains
are rougher than the small grains as in the experiments and in the thick
flow regime above. However, we find that the transition to segregation
does not occur sharply at $\delta = 0$ but occurs at a small negative
value $\delta_c<0$, which depends on the value of the
cross-amplification rates $\gamma_{\alpha\beta}$. When the cross-rates
are zero, then the transition occurs at $\delta = 0$ as we found for the
thick flow regime above. For smaller values of $\delta<\delta_c<0$ we
find the complete segregation pattern found in the experiments. Thus,
the presence of cross-amplification processes---which appear only in the
low flux limit---shifts the transition from stratification to
segregation to a value $\delta_c$ different from zero \cite{baxter}.
However, the stratification we find for $\delta_c<\delta<0$ is less
pronounced than in the case $\delta>0$: the layers do not go to the top
of the sandpile, and the layer of small grains is very thin. Similar
results at the neighborhood of the stratification-segregation transition
have been found with a microscopic model of grains dynamics \cite{mh}.
We also find a kink, corresponding to the growth of the new pair of
static layers, with a well-defined steady-state profile and upward
velocity for $\delta >0$.

This model, valid for thin flows, is qualitatively close the preceding
one valid for thick flows, 
where we also found a regime of 
 complete segregation separating the
surface of the sandpile in two regions. This can be understood noting
that the dependence of the repose angle on the composition of the
surface in the thin flow model is comparable to the effect of
percolation in the rolling layer in the thick flow regime. 
Since for a given composition
$\phi_\beta$, the repose angle of the small grains is always larger than
the repose angle of the large grains (Fig.~\ref{contia}), the small
grains are also the first to be trapped before the large ones. The
composition of small grains on 
the surface increases, amplifying the
effect, up to the point where the surface is made of only small grains.

\section{Analytical results}

The thick flow and the thin flow regime show similar results as long as
there is a large difference in size between the grains. We will now show
analytical results valid for both regimes.

\subsection{Steady state solution}
\label{steady}

We next study the stationary solution of Eqs.~(\ref{eq:R_et_h}) and then
we study how a perturbation to this solution can be amplified giving
rise to stratification. For simplicity, we consider the geometry of a
silo of lateral size $L$, i.e., all the rolling grains are stopped when
they reach a wall at position $x=L$ [Fig.~\ref{cell}(a)].  Moreover,
we assume the difference
\begin{equation}
\psi \equiv \theta_2(\phi_2)-\theta_1(\phi_2)
\end{equation}
to be independent of the concentration $\phi_2$. We seek for a solution
where the profiles of the sandpile and of the rolling grains are
conserved in time. Thus, stratification that corresponds to periodic
variations in time of the different variables of interest cannot be
observed for this solution. The conservation of the grains gives
\cite{drop}
\begin{equation}
\frac{\partial h}{\partial t} = \frac{v ~ R^0}{L},
\end{equation}
and we impose
\begin{equation}
\frac{\partial R_\alpha}{ \partial t}(x) = 0.
\end{equation}
 We assume that, as in the experiment, an equal volume mixture is
used, so
\begin{equation}
R_1^0 = R_2^0 = R^0/2,
\end{equation} 
and the boundary conditions are
\begin{equation}
\begin{array}{l}
R_\alpha (L) =0,\\ R_\alpha (0) =R^0/2.
\end{array}
\end{equation}

The steady-state solution of (\ref{eq:R_et_h})-(\ref{canonical1})
shows {\it total} segregation: except in a region of size of
$v/\gamma$ at the center of the sandpile where the grains are mixed,
the sandpile is exclusively made of small grains in the upper part of
the pile, and of large grains in the lower part of the pile. The
details of the calculations can be found in \cite{makse3}. Here we
present a simplified results for the case of $v/\gamma \ll L$, since
we expect (\ref{voverg}).

At the {\it upper\/} part of the pile, for $0\le x\le x_m$, with $x_m \equiv
L/2-v/(\gamma \psi)$, only small grains are present ($\phi_1(x)=1$,
and $\phi_2(x) = 0$), and the profiles are
\begin{mathletters}
\label{steady-up}
\begin{eqnarray}
R_1(x) &=& R^0 \left(\frac{1}{2} -\frac{x}{L}\right), \\ R_2(x) & = &
R^0/2, \\ \theta(x)-\theta_{11} & =& - \frac{v/\gamma+\psi\gamma_{21} 
L/(2\gamma)}{ L (1+\gamma_{21}/\gamma)/2-x}.
\end{eqnarray}
\end{mathletters}
At the {\it lower\/} part of the pile ($x_m \le x \le L$), we find that
after a small region of size of the order of $v/(\gamma \psi)$ (or
equivalently, of the order of the size of the grain), mainly large
grains are present. The profiles are
\begin{mathletters}
\begin{eqnarray}
\phi_1(x) & = & \exp\left[{-\frac{\gamma\psi}{v}(x-x_m)}\right], \\
R_1(x)& =& \frac{2v}{\gamma \psi L}\phi_1(x) R (x), \\
\theta(x)-\theta_{22} &=& -m\phi_1(x)-\frac{v}{\gamma (L-x)}.
\end{eqnarray}
\label{steady-down}
\end{mathletters}
Here
\begin{equation}
R(x)\equiv R_1(x)+R_2(x)=R^0(1-x/L),
\end{equation}
and $m\equiv \theta_{22}-\theta_{21}=\theta_{12}-\theta_{11}$.

When the system is in this stationary state, the angle of the upper
part of the sandpile (where only small grains exist) is almost
$\theta_{11}$; without cross-amplification ($\gamma_{21}=0$), 
the sandpile
would be at its repose angle $\theta_{11}$, and because of
cross-amplification, the angle is significantly reduced. The
divergence around $x=L/2$ insures that all the small grains are
captured despite cross amplification. Moreover the angle for the low
part of the sandpile is almost $\theta_{22}$. If we now suppose a
fluctuation in $R_1$ such that few small rolling grains enter the low
part of the sandpile, capture will be of the order of $\gamma
(\theta_{22}-\theta_1(\phi_2))R_1$, and amplification will be of the
order of $\gamma (\theta_{22}-\theta_2(\phi_2))R_2$. As $R_1\ll R_2$,
amplification will dominate capture, and the small rolling grains will
go downhill up to the point when capture starts to exceed
amplification. As soon as this small layer of rolling grains exist,
the large grains are convected down without being capture. This excess
of large grain creates a kink at the bottom of the pile going uphill
where large grains are now stopped; the first layers are formed and
the same process starts again. This may explain why when
cross-amplification terms are present the transition
stratification-segregation is shifted from $\delta=0$ to $\delta =
\delta_c<0$. 
\begin{figure}[htbp]
\centerline{ \vbox{ \hbox{\epsfxsize=8cm
\epsfbox{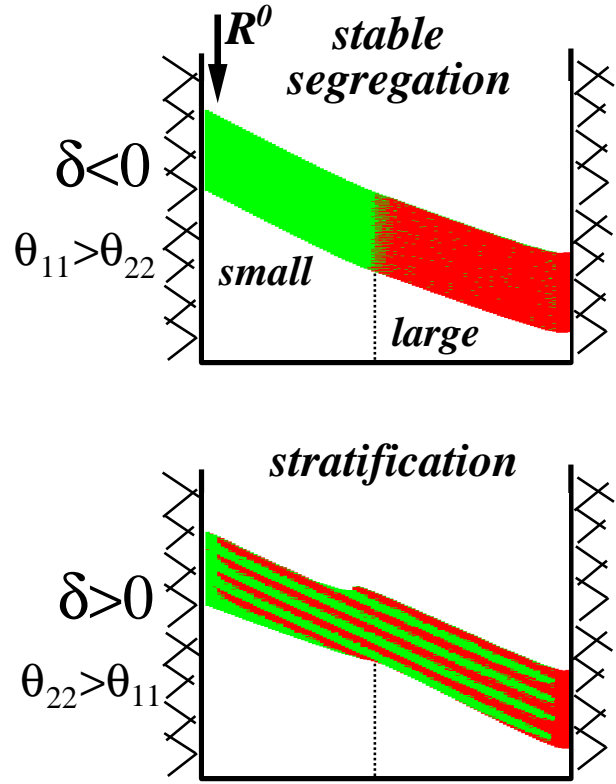} } }
\vspace{.5cm}} 
\narrowtext
\caption{Stability analysis of the steady state solution of the
equations of motion for granular flow of mixtures in the geometry of
the silo. When $\delta<0$ we find that the steady-state solution is
stable, so that this solution is the physical solution when
$\delta<0$. On the other hand, when $\delta>0$ the steady state
solution is unstable and it evolves into stratification as in the
experiment.}
\label{stability}
\end{figure}
When percolation is present this shift disappears since
cross-interactions are greatly reduced by the screening effect of the
small rolling grains at the bottom of the rolling phase.

To analyze the stability of the steady-state solution for the different
phenomenologic parameters, we impose the steady-state solution as the
initial condition, and then we look numerically for the stability of the
profile under perturbations (Fig.~\ref{stability}). For
$\theta_{11}>\theta_{22}$, the steady state solution is stable, so that 
Eqs. (\ref{steady-up}) and  (\ref{steady-down}) are the physical solution
for this case. In this
case, only segregation is observed, and the sandpile conserves in time
the profiles (\ref{steady-up}) and (\ref{steady-down}). For
$\theta_{11}<\theta_{22}$, the steady-state solution is {\it unstable\/}
(evolving to stratification), just as in \cite{makse}. The onset of the
instability is clearly seen, where the kink first start at the center of
the pile where some large grains are captured on top of small grains as
seen in Fig.~\ref{stability} for $\theta_{22}>\theta_{11}$.

Our model predicts the complete segregation of the species when 
$\theta_{22}<\theta_{11}$. The concentration of small grains drops very fast
near the center of the pile in a characteristic region 
of the order of 
$v / (\gamma\psi) \sim $cm according to experiments \cite{makse2}.
As discussed in Sec. \ref{collision} this is a consequence of our assumption 
of large difference in properties of the species.

\subsection{Analytical Shape of the Kink}

We saw in a preceding section that the layers were constructed through
a kink where all the grains are stopped. A precise analysis of this
kink would allow us to understand the dynamic of the process and the
characteristic size of the layers.

From the simulations
we observe that the kink has a well
define uphill speed and that its shape is preserved in time. To obtain
an analytically tractable set of equations, we then look for a
possible steady-state solution for the shape of the kink, and thus for
$R(x,t)$ and $h(x,t)$. We also must make the assumptions that far
below and above the kink, the sandpile has a constant angle $\theta_0$
and is only made of large grains. These assumptions are verified by
comparing the obtained result for the shape of the kink with the
numerical results.

The existence of a stationary solution for the kink implies
that $R_\alpha (x,t)$ and $f(x,t)\equiv h(x,t)+\theta_0 x$ are
functions only of
\begin{equation}
u\equiv x+v_\uparrow t,
\end{equation}
where $v_\uparrow$ is the uphill speed of the kink. For the lowest
layer of the kink composed of small grains (the upper part of the pile), 
as only small grains are captured ($\phi_1(u)=1$,
$R_2(u)=R^0/2$), Eqs. (\ref{eq:R_et_h}) reduce to equations for
$R_1(u)$ and $f(u)$ (we assume $\gamma_{21}=\gamma$), 
\begin{equation}
\begin{array}{rll}
{\displaystyle (v_\uparrow + v )\frac{\partial R_1(u)}{\partial
u}}&=&{\displaystyle \gamma \left ( -\frac{\partial f}{\partial
u}-\delta_{11} \right ) R_1 + }\\ & &{\displaystyle \gamma \left (
-\frac{\partial f}{\partial u}-\delta_{21} \right ) R_2^0 } \\
{\displaystyle v_\uparrow \frac{\partial f(u)}{\partial u}}
&=&{\displaystyle \gamma \left ( \frac{\partial f}{\partial
u}+\delta_{11} \right ) R_1 } + \\ & &{\displaystyle \gamma \left (
\frac{\partial f}{\partial u}+\delta_{21} \right ) R_2^0},
\end{array}
\end{equation}
with
\begin{equation}
\delta_{11} \equiv \theta_{11} - \theta_0, ~~ \delta_{21} \equiv
\theta_{21} - \theta_0
\end{equation}

We obtain the shape of the low part of the kink. For $u \le 0$,
$f(u)=0$ and for $u>0$, $f(u)$ and $R_1(u)$  obey 
the following equations:

\begin{equation}
\label{eq:kinkComplete}
\begin{array}{ccl}
{\displaystyle \frac{\partial f}{\partial u}}&=&
{\displaystyle \frac{(\delta_{11} + \delta_{21}) R^0/2 - w
\delta_{11} f(u)}{ v_\uparrow/\gamma +w f(u) - R^0},}\\
R_1(u) &=& - w f(u) + R^0/2
\end{array}
\end{equation}
where $w \equiv v_\uparrow/(v+v_\uparrow)$. Then the lower part of
the kink is characterized by a linear dependence
\begin{equation}
\label{eq:startKink}
f(u) \propto \frac{(\delta_{11}+\delta_{21})/2}{v_\uparrow/(\gamma R^0)-1} u.
\end{equation}
Finally, this solution is valid up to the point where the large grains
start to be captured, i.e., $\partial f / \partial u =-\delta_{21}$.
From Eqs.~(\ref{eq:kinkComplete}) and (\ref{eq:startKink}) we obtain
the following inequalities:
\begin{equation}
\theta_0> \frac{\theta_{11}+\theta_{21}}{2}, ~~ v_\uparrow
<\gamma R^0.
\end{equation}
The solution of the equations for the upper layer of the kink (the  lowest 
part of the pile) where
only large grains are present can be obtained in the same way and is
\begin{equation}
\label{upper_kink}
f(u) = \left({R^0\over w}\right) \exp \left({w\gamma\delta_{22} u\over
   v_\uparrow}\right),
\end{equation}
where
\begin{equation}
\delta_{22}\equiv\theta_0-\theta_{22}.
\end{equation}
We then find that the shape of the upper part of the kink is
exponential, and that the stationary solution exists only for
$\delta_{22}<0$.

Thus we see that the existence of the stationary solution for the kink
implies that
\begin{equation}
\frac{\theta_{11}+\theta_{21}}{2}<\theta_{22},
\label{eq:condStrat}
\end{equation}
and that the sandpile is built on an intermediate angle $\theta_0$
between those
two extreme values. We are now able to distinguish two types of
stratification. When Eq.~ (\ref{eq:condStrat}) is satisfied, a
stationary solution exists for the kink and the layers are well
formed. When Eq.~ (\ref{eq:condStrat}) is not satisfied, the layers
are very asymmetric, with thin layers of small grains.
 
\subsection{Wavelength of the Layers} 

The layer thickness $\lambda$ is defined by the width of the kink,
$\lambda \equiv \lim_{u\rightarrow \infty}f(u)$. From
Eq.~(\ref{upper_kink}), we obtain $\lambda = R^0/w,$ which is a
consequence of the conservation law stating that all the rolling grains
are stopped at the kink. This relation is obtained assuming that the
density of the fluid phase is the same as the density of the bulk.
However, in general we have $\mu_{\mbox{\scriptsize fluid}} <
\mu_{\mbox{\scriptsize bulk}}$, where $\mu_{\mbox{\scriptsize fluid}}$
and $\mu_{\mbox{\scriptsize bulk}}$ are the number of rolling grains
per unit volume of the fluid phase and bulk, respectively. Then we
obtain
\begin{equation}
\lambda = \frac{\mu_{\mbox{\scriptsize fluid}}}
{\mu_{\mbox{\scriptsize bulk}}} ~\frac{(v+v_\uparrow)} { v_\uparrow}~
R^0.
\label{conser}
\end{equation}
Furthermore, Eq.~(\ref{eq:startKink}) implies that
\begin{equation}
v_\uparrow = C_\uparrow \gamma R^0,
\end{equation}
where $C_\uparrow$ is a numerical constant that does not depend on
$v$, $\gamma$, or $R^0$. Then we obtain
\begin{equation}
\label{lambda1}
\lambda_{\mbox{\scriptsize thin}} ~= \frac{\mu_{\mbox{\scriptsize fluid}}}
{\mu_{\mbox{\scriptsize bulk}}} \left ( \frac{v}{C_\uparrow\gamma} ~+~
R^0 \right ) \qquad \mbox{[thin flow regime].}
\end{equation}

This relation is relevant to the thin flow regime where $v/\gamma\sim
d$ and $R^0\sim d$ are both of the order of the size of the grains,
and the velocity of the rolling grains is constant since the rolling
phase is homogeneous. However in the thick flow regime the mean value
of the velocity of the grains scale with the thickness of the rolling
phase as found recently experimentally \cite{makse2}
\begin{equation}
v = C ~\gamma ~R^0,
\end{equation}
where $C$ is a numerical constant that does not depend on $\gamma$ or
$R^0$, but may depend on the angles of repose and other features of
the grains. In this case Eq. (\ref{conser}) becomes
\begin{equation}
\lambda_{\mbox{\scriptsize thick}} = \frac{\mu_{\mbox{\scriptsize fluid}}}
{\mu_{\mbox{\scriptsize bulk}}} ~\left (1+\frac{C}{C_\uparrow}\right )
~R^0 \qquad \mbox{[thick flow regime].}
\label{lambda2}
\end{equation}

Thus we see that we expect a linear dependence of the wavelength of
the layers as a function of the thickness of the rolling phase, i.e.
as a function of the flux of grains (if the plate separation is
maintained constant \cite{makse2}), but the proportionality constant
is different in the thin flow regime and the thick flow regime.
From (\ref{lambda1}) and (\ref{lambda2}) we can obtain an approximate 
estimation of the value of 
the cross-over from the thin flow regime to the thick flow
regime 
$R^0_c$ 
assuming that  $\lambda_{\mbox{\scriptsize thin}}(R_c^0) \approx
\lambda_{\mbox{\scriptsize thick}}(R_c^0) $, then
$R_c^0 \approx v_{\mbox{\scriptsize thin}}/ (C \gamma),$
where $v_{\mbox{\scriptsize thin}}$ is the velocity of the rolling
grains in the thin flow regime.

Typical
experimental values of the phenomenological constants of the problem 
for a system of $L=30$ cm, $d_1=0.27$ mm, $d_2=0.8$ mm, 
are
\cite{makse3} 
$v\approx 10$ cm/sec, $\gamma \approx 20$/sec,
$\tan\psi=\tan \theta_{11} - \tan \theta_{21} \approx 0.2$,
$\mu_{\mbox{\scriptsize fluid}}/ \mu_{\mbox{\scriptsize bulk}} =
0.85$, and $C/C_\uparrow\approx 1$. Therefore, $v/(\gamma \tan\psi)
\simeq 2.5$ cm. 
We also obtain $R_c^0 \approx 0.5$ cm as the value of
the cross-over flux from the thin flow regime to the percolation
regime. 
The above predictions regarding the dependence of the
wavelength on the flux of grains 
have been confirmed in recent
experiments \cite{makse2,yan}.

\section{Discussion}
\label{discussion}

In summary, we develop a mechanism to explain the spontaneous
stratification observed in \cite{makse}. This mechanism is related to
the dependence of the local repose angle on the local surface
composition. When we consider only capture (corresponding to the
cellular automaton model), we find that spontaneous stratification
occurs only when the repose angle of the large grains is larger than
the repose angle of the small grains ($\theta_{22}>\theta_{11}$,
corresponding to large grains rougher than small grains). This result
is in agreement with the experiments. Stratification is also obtained
when $\theta_{22}$ is slightly smaller than $\theta_{11}$ as soon as
we include cross-amplification process, as we find in the thin flow 
regime with the continuum model.
When $\theta_{22}<\theta_{11}$, the model predicts almost complete
segregation, but not stratification. These results are in agreement
with experiments \cite{makse}.

It is interesting to compare those results with a previous study
\cite{makse} where a continuum model was used neglecting
cross-amplification (i.e., amplification of $\alpha$ static grains by
$\beta$ rolling grains). In this case, it was shown that amplification
never occurs during the formation of the sandpile, and stratification
is obtained in the same conditions as in the experiments.

We discuss the possible importance of percolation effects in the
rolling layer, that takes place when the input flux of grains is large.
Although the continuum model must be deeply modified, the results
are surprisingly close to the thin flow model.

The model describes well the static picture of the sandpile of
\cite{makse} with alternating layers made of small and large grains,
and also reproduces the dynamics, where the layers are built through a
kink mechanism. The numerical simulations suggest that the motion of
the kink is stationary. It allows us to make quantitative predictions
for the dependence of the size of the layers on the different
parameters of the problem.

From the theoretical point of view, it is the first time that this
continuum formalism for granular flow is directly compared to
experiments. The quality of the results suggest that this formalism
includes the essential features of the physics of granular flow.  Using
this formalism, it may be possible to have theoretically access to 3D
problems, such as periodic segregation in rotating cylinders
\cite{zik}, that are out of reach with present-day computer simulations.

\subsubsection*{Acknowledgments}
We thank T. Boutreux, P. G. de Gennes, S. Havlin, H. J. Herrmann, P. R.
King, and S. Tomassone for stimulating discussions, and NSF and BP for
financial support.





\end{multicols}

\begin{thebibliography}{99}
\bibitem{time}
In this case, very small time steps in the numerical integration
of the equation
of motion
must be used to avoid instabilities which leads to  oscillations 
in the center of the pile.
\bibitem{drop} We drop the time dependence of the variables in the
steady state.
\end{thebibliography}
\end{document}